\documentclass{aa}
\usepackage{graphics,epsf,epsfig,graphicx}
\usepackage{natbib}
\usepackage{url}
\usepackage{pstricks,psfrag}
\usepackage{amssymb}

\bibpunct{(}{)}{;}{a}{}{,} % to follow the A&A style

\newcommand{\be}{\begin{equation}}
\newcommand{\ee}{\end{equation}}
\newcommand{\dd}{\mathrm{d}}

\begin{document}

\title{Probing dust grain evolution in IM Lupi's circumstellar disc}
\subtitle{Multi-wavelength observations and modelling of the dust disc}

\author{C.~Pinte\inst{1,2},
  D.L.~Padgett\inst{3},
  F.~M\'enard\inst{2},
  K.R.~Stapelfeldt\inst{4},
  G.~Schneider\inst{5}, 
  J.~Olofsson\inst{2},
  O.~Pani\'c\,\inst{6}, 
  J.C.~Augereau\inst{2}, 
  G.~Duch\^ene\inst{2,7}, 
  J.~Krist\inst{4},
  K.~Pontoppidan\inst{8}, 
  M.D.~Perrin\inst{9},
  C.A.~Grady\inst{10}, 
  J.~Kessler-Silacci\inst{11},
  E.F.~van Dishoeck\inst{6,12},
  D.~Lommen\inst{6},
  M.~Silverstone\inst{13},
  D.C.~Hines\inst{14},
  S.~Wolf\inst{15},
  G.A.~Blake\inst{8},
  T.~Henning\inst{16}, 
  B.~Stecklum\inst{17}.
}

\offprints{C. Pinte \\ \email{pinte@astro.ex.ac.uk}}
\institute{
School of Physics, University of Exeter, Stocker Road, Exeter EX4 4QL, United Kingdom
\and 
Laboratoire d'Astrophysique de Grenoble, CNRS/UJF UMR~5571, 
414 rue de la Piscine, B.P. 53, F-38041 Grenoble Cedex 9, France
\and
Spitzer Science Center, Caltech, Pasadena, CA 91125, USA
\and
Jet Propulsion Laboratory, California Institute of Technology,
Pasadena, CA 91109, USA
\and
Steward Observatory, The University of Arizona, 933 North Cherry Avenue, 
Tucson, AZ 85721, USA
\and
Leiden Observatory, Leiden University, P.O. Box 9513, 2300 RA Leiden,
The Netherlands
\and
Astronomy Dept, UC Berkeley, Berkeley CA 94720-3411, USA
\and
Division of Geological and Planetary Sciences 150-21, California
Institute of Technology, Pasadena, CA 91125, USA
\and
Department of Physics and Astronomy, UCLA, Los Angeles, CA 90095-1562, USA 
\and
Eureka Scientific and Goddard Space Flight Center, Code 667,
Greenbelt, MD~20771, USA 
\and
The University of Texas at Austin, Department of Astronomy, 1
University Station C1400, Austin, Texas 78712--0259, USA
\and
Max Planck Institut f\"ur Extraterrestrische Physik,
Giessenbachstrasse 1, 85748 Garching, Germany
\and
Eureka Scientific, Inc., NC Branch, 113 Castlefern Dr., Cary, NC 27513
\and 
Space Science Institute, Corrales NM, 87048, USA 
\and
University of Kiel, Institute of Theoretical Physics and Astrophysics,
Leibnizstrasse 15, 24098 Kiel, Germany
\and
 Max Planck Institute for Astronomy, K\"onigstuhl 17, 69117
 Heidelberg, Germany
\and
 Th\"uringer Landessternwarte Tautenburg, Sternwarte 5, 07778
 Tautenburg, Germany 
}

\authorrunning{C. Pinte et al.}
\titlerunning{}

\date{Received ... / Accepted ...}

\abstract{}{We present a panchromatic
  study,  involving a multiple technique approach, of the circumstellar disc surrounding the T~Tauri star IM~Lupi
  (Sz~82).}
{We have undertaken a comprehensive observational study of
  IM Lupi using photometry, spectroscopy, millimetre interferometry
  and multi-wavelength imaging.
For the first time, the disc is resolved from optical and near-infrared wavelengths in
scattered light, to the millimetre 
regime in thermal emission.
Our data-set, in conjunction with existing photometric
  data,  provides an extensive coverage of the spectral energy
distribution, including a detailed spectrum of the silicate emission bands.
We have performed a simultaneous modelling of the various
observations, using the radiative transfer code MCFOST, and analysed a
grid of models over a large fraction of the parameter space
via Bayesian inference.
}{We have constructed a model that can reproduce all of the
  observations of the disc. Our analysis illustrates the importance of
  combining a wide range of 
observations in order to fully constrain the disc model, with each
observation  providing a strong constraint only on some aspects of the
disc structure and dust content. 
Quantitative evidence of dust evolution in the disc is obtained:
  grain growth up to millimetre-sized particles, vertical stratification of dust
  grains with micrometric grains close to the disc surface and larger
  grains which have settled towards the disc
  midplane, and possibly the formation of fluffy aggregates and/or ice
  mantles around grains.}{}

\maketitle

%======================================================================
\section{Introduction}
During the first stages of planet formation following the core nucleated accretion
scenario \citep{Lissauer06PPV}, evolution of dust grains within the
protoplanetary disc surrounding the central forming 
object is expected. Micrometre size dust grains will start to grow by coagulation 
during low relative velocity collisions, leading to the formation of
larger, potentially fluffy 
aggregates \citep[][and references
therein]{Beckwith00,Dominik06PPV,Natta06PPV}
 that will ultimately give birth to kilometre size
planetesimals promoting gravitational focusing and eventually leading to
planets. Timescales for the formation of these
aggregates strongly depend on the physics of aggregation as well as the physical
conditions 
within the disc, such as differential velocities between grains and grain-gas
interactions. 
  
\begin{figure*}
  \includegraphics[height=0.32\hsize]{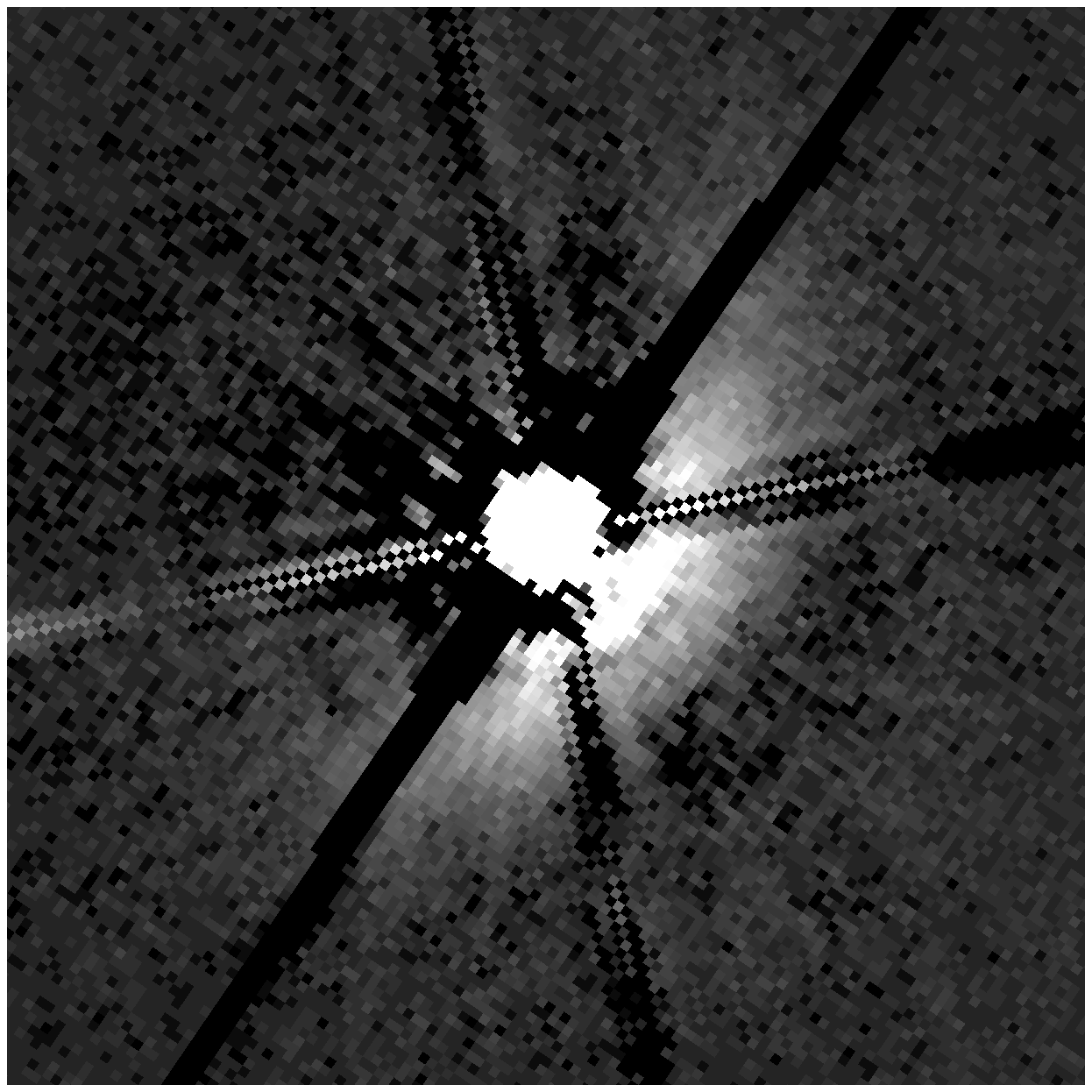}%
  \hspace{\stretch{0.5}}%
  \includegraphics[height=0.32\hsize]{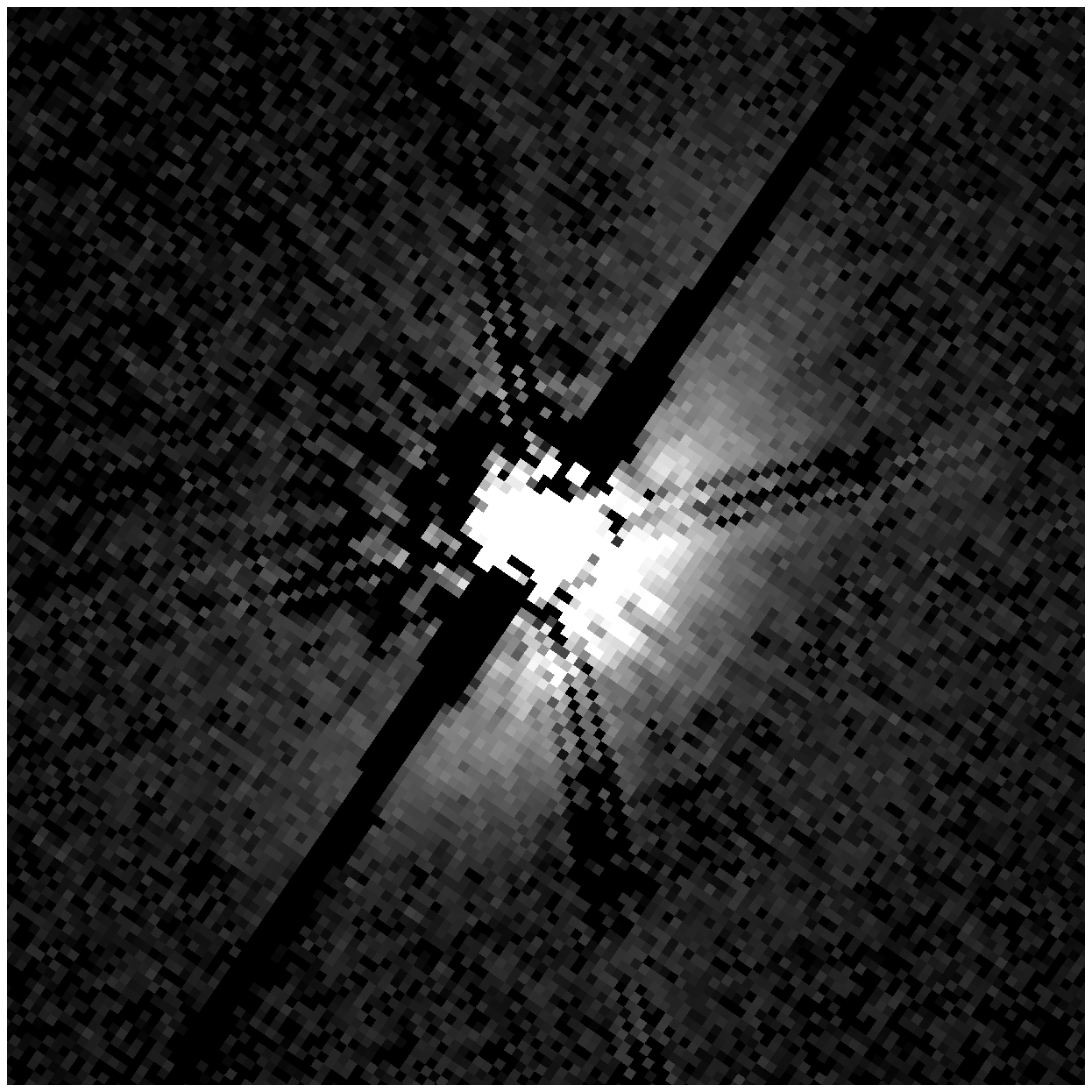}%
  \hspace{\stretch{0.5}}%
  \includegraphics[height=0.32\hsize]{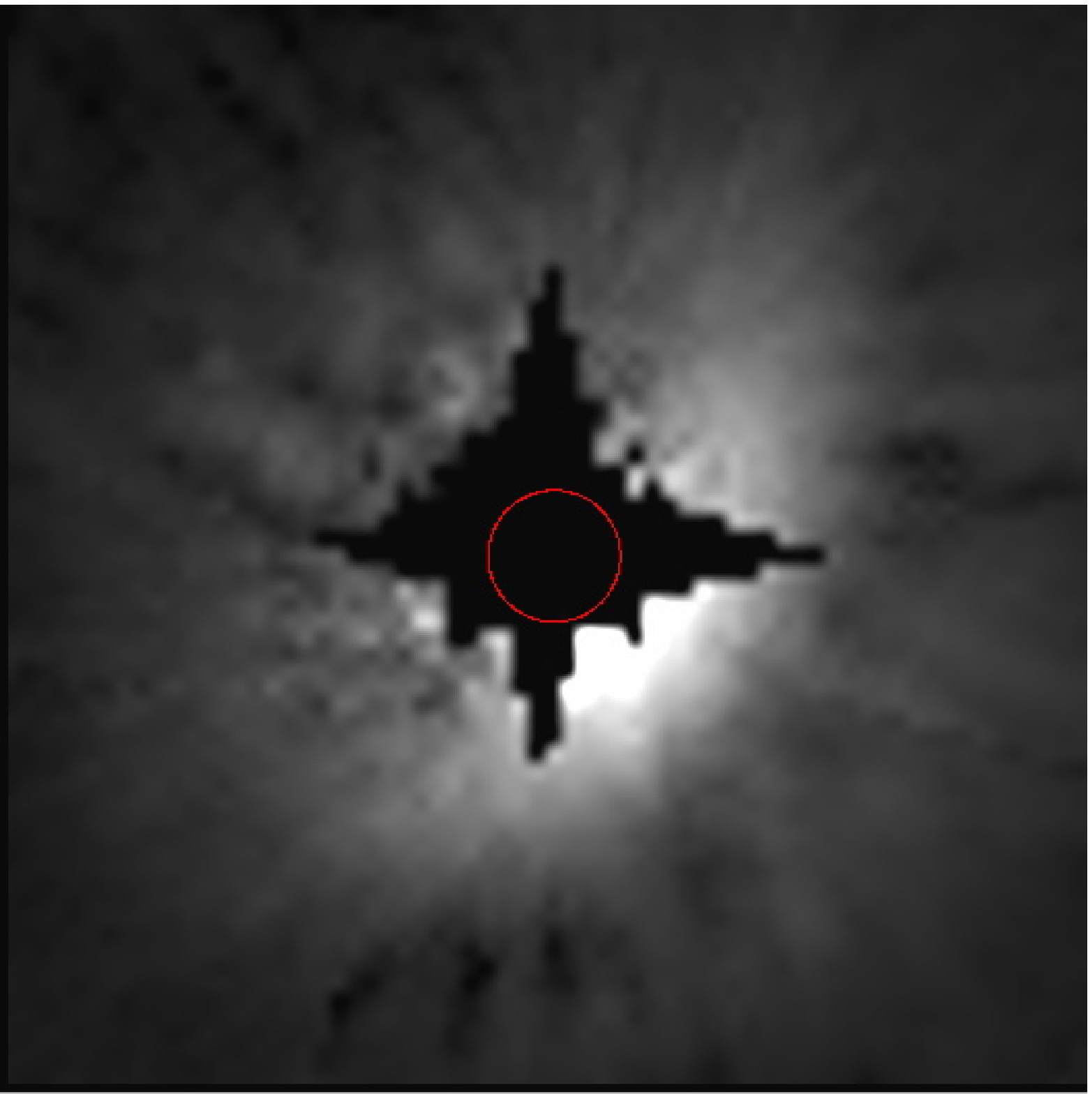}%
  \caption{ The IM Lupi circumstellar disc observed in scattered light
    by HST.
    {\bf Left and middle panels:} respectively F606W and F814 \emph{WFPC2}
    PC1 PSF-subtracted images.  The dark diagonal feature running through the star is an
    artifact of charge bleeding along the CCD detector columns. 
    {\bf Right panel:} \emph{NICMOS} PSF-subtracted coronagraphy at 1.6 $\mu$m. 
    The central
    circle represents the 0.3\arcsec\ radius coronagraphic
    obscuration. All images are shown in square root stretch, with
    North up and East to the left.  The field of view is 5x5\arcsec
    (dashed boxed in Fig.~\ref{fig:schema}). 
    \label{fig:obs_HST}}
\end{figure*}

\begin{figure}
  \centering
  \includegraphics[width=\hsize]{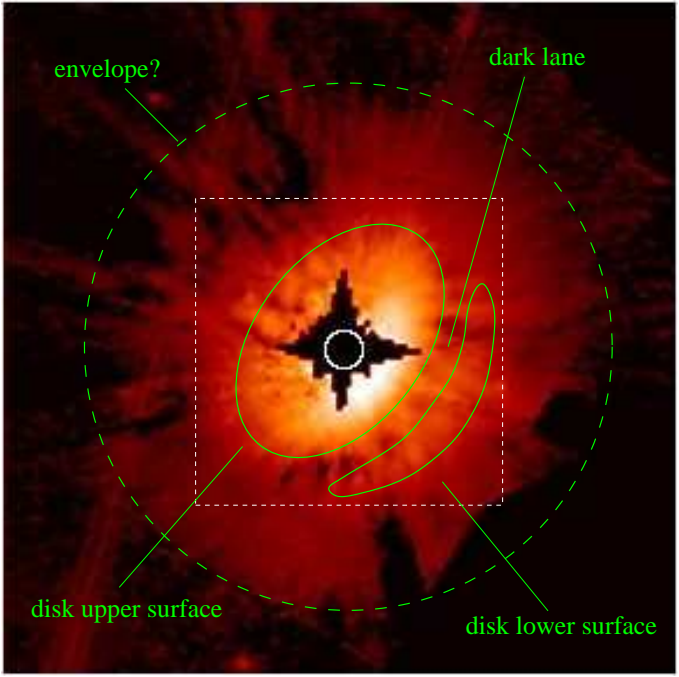}
  \caption{\emph{NICMOS} F160W image in $\log$ stretch. The field of view
    is 11$\times$11\arcsec\ with North up and East to the left. The
    dashed box corresponds to the field of view in
    Fig.~\ref{fig:obs_HST}. The central white circle represents the
    0.3\arcsec\ radius coronagraphic obscuration. The full green lines
    indicate the upper and lower scattering surfaces of the disc
    (separated by the dark lane which corresponds to the disc midplane). The
    green dashed circle represents the
    possible envelope surrounding the disc.\label{fig:schema}}
\end{figure}

In parallel to grain growth, dust grains are
expected to settle towards the disc midplane and then to migrate
inward as a result of the conjugate
actions of the stellar gravity and gas drag \citep[e.g.][]{Barriere05,Fromang06}.
 This settling is highly dependent on the grain size. Small grains ($<$~1~$\mu$m)
 are strongly coupled to the gas, they follow its motion, and do not settle
 at all. Conversely, grains in the mm-cm regime are considerably slowed
 down by the gas drag. They completely
 decouple and settle very rapidly in a thin midplane.

The result is the formation of a dust sub-disc close to the equatorial
plane \citep[e.g.,][]{Safronov69,Dubrulle95} that may become unstable and
produce planetesimals when the density of the dust layer exceeds the
gas one \citep[e.g.,][]{Goldreich73,Schrapler04,Johansen07}. 
Models suggest grain growth to large
particle sizes with a population of small grains remaining close to the
surface and larger grains deeper in the disc \citep[e.g.][]{Dominik06PPV}.
However, many details of process remain uncertain. For example, it is
not clear 
how dust grains overcome the ``metre size barrier''
without being accreted onto the star \citep[e.g.\
][]{Weidenschilling77,Brauer08} or destroyed by high speed collisions
\citep[e.g.][]{Jones96,Blum00}, or how Kelvin-Helmholtz instabilities prevent the
dust sublayer from fragmenting \citep[e.g.][]{Johansen06}.

\defcitealias{Dullemond04}{D04}
\def\D04{\citetalias{Dullemond04}}

Detection of the large building blocks of planets (particle sizes $> 1$m) in the disc midplane
is far beyond current observational capacities, but we can expect to
detect signatures of their formation. 
The stratified structure resulting from grain growth and
settling has direct consequences on disc
observables like their spectral energy distributions (SEDs) and
scattered light images \citep[e.g.][hereafter D04]{Dullemond04}.
A variety of observational techniques have been used to obtain insights into the disc
properties and their dust content:  SEDs
\citep[e.g.][]{D'Alessio01}, scattered light images \citep{Watson06PPV}, thermal emission maps in
the millimetre regime \citep{Dutrey06PPV}, mid-infrared spectroscopy \citep[e.g.][]{Kessler-Silacci06}.
However, each technique only provides a limited view of a disc.
SED analysis leads to multiple ambiguities (e.g.,
\citealp{Thamm94,Chiang01}) since the lack of spatial resolution
precludes from
solving the degeneracies between model parameters like geometry and opacity. 
Spatially resolved observations in a single spectral band (scattered light in the optical or
near-infrared regime or thermal emission in the millimetre domain) also
 gives incomplete information about the disc.
As the dust opacity is a steep function of wavelength and because high
temperature  gradients are present
within the disc, a single-band observation gives insight to only a limited
region of the disc. Thus, scattered light images probe the surface
layers of these optically thick discs at large radii whereas, millimetre observations are mainly
sensitive to the bulk of the disc mass closer to the midplane. 
To precisely study fine physical processes like dust
evolution and stratification, it is necessary to combine the
aforementioned methods in a
multi-wavelength, multiple technique
observational and modelling approach.

In this paper, we study the circumstellar environment of the T~Tauri
star IM~Lupi.   The disc surrounding IM~Lupi  (Schwartz~82, HBC~605,
  IRAS~15528-3747) was
first detected in scattered light
with the {\it WFPC2} instrument on board the \emph{Hubble Space Telescope}
(\emph{HST}) as part of a T~Tauri star
imaging survey (Stapelfeldt et al. 2008, in prep.). IM~Lupi  is 
an M0 T~Tauri star located within the Lupus star forming 
clouds. It is one of four young stellar objects in the small 
$^{13}$CO(1-0) Lupus~2 core near the extreme T~Tauri star RU~Lupi \citep{Tachihara96}.
Despite the low accretion-related spectroscopic activity of
IM~Lupi \citep{Reipurth96,Whichmann99}, longer wavelength observations reveal ample  
evidence for circumstellar material in the system with strong mm continuum
emission \citep{Nuernberger97}.  Single dish $^{12}$CO and $^{13}$CO
line observations indicate that the disc is gas rich and are  
consistent with a rotating disc model \citep{vanKempen07}. The
3.3\,mm continuum emission from  the disc was spatially resolved by
\cite{Lommen07}. Preliminary models by \cite{Padgett06}
showed that the infrared excess is well-reproduced by a disc model.

We have built a rich data set of observations of the
disc: a well sampled SED,
\emph{HST} multiple
wavelength scattered light images, \emph{Spitzer} near- and mid-infrared spectroscopy
and \emph{SMA} millimetre emission maps. These observations provide
different, complementary detailed views of the disc structure and dust properties.
We investigate whether all of these
observations can be interpreted in the framework of a single model
from optical to millimetre wavelengths, and
whether we can derive quantitative conclusions regarding the evolutionary stage of
the disc.  In section~\ref{sec:obs}, we describe the observations and data reduction
procedure. In
section~\ref{sec:simple_estimation}, we first draw constraints on the
disc and dust properties from 
the various observations and then, in
section~\ref{sec:grid_modelling}, we present simultaneous modelling of
these observations.
A detailed study of the dust properties is presented in
section~\ref{sec:dust_properties}. In section~\ref{sec:discussion},  we
discuss the implication of the results, and 
section~\ref{sec:conclusion} contains the concluding remarks.

%======================================================================
\section{Observations}
\label{sec:obs}

\begin{figure}
  \includegraphics[width=\hsize]{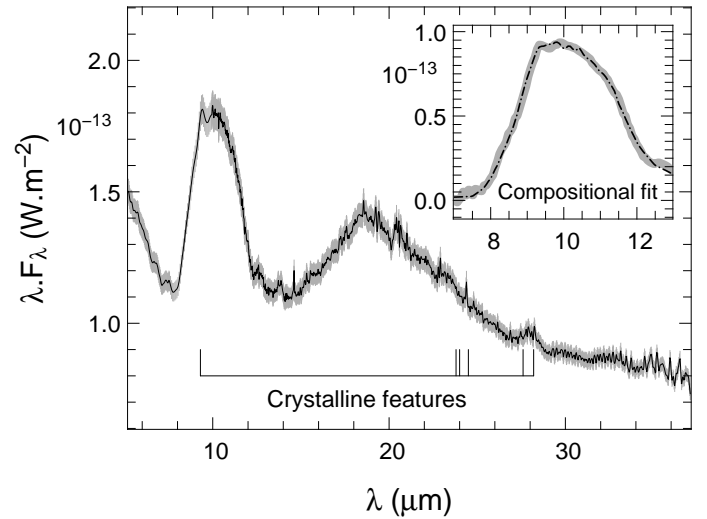}
  \caption{\emph{Spitzer/IRS} spectrum of IM Lupi. The large
    features of amorphous silicates are seen at 10 and 18\,$\mu$m, as well as
    crystalline features at 9.3\,$\mu$m and around 27\,$\mu$m. A blend
    around 24$\,\mu$m is tentatively detected.
    The
    spectrum has been slightly smoothed to reduced the noise. The
    jump at 20\,$\mu$m is an instrumental artifact resulting from the
    transition between the Long-Low and Long-High modules of
    \emph{IRS}. The inset shows the fit of the 10$\,\mu$m continuum subtracted silicates
    feature. The thick grey line represents the
    observed spectrum (corresponding to the black line in the main panel) and the dot-dashed black line is the fitted synthetic
    spectrum (see section \ref{sec:mineralogie}). \label{fig:silicates}}
\end{figure}

\subsection{Scattered light images}

\subsubsection{HST/WFPC2  observations and data processing}

{\it HST} imaging observations of the
protoplanetary disc surrounding IM~Lupi were obtained as part of an {\it WFPC2} snapshot 
survey of nearby T~Tauri stars ({\it HST} Cycle 7 GO/7387 program, PI: Stapelfeldt),
on 1999 February 18, using the HST Planetary Camera~1 (spatial scale
of 45.5\,mas.pixel$^{-1}$). The data 
consist of short and long exposures through the F606W and 
F814W filters (F606W: 8 s and 100 s; F814W: 7s and 80 s) which 
roughly correspond to Johnson $R$ and $I$.
The spatial resolution in both filter
bands F606W and F814W is 0.06\arcsec and 0.08\arcsec, respectively.  
In the long exposures, the star is heavily saturated in an attempt to
reveal low surface  
brightness circumstellar nebulosity. After standard \emph{WFPC2} 
data reduction, the stellar point spread function (PSF) was 
compared to our large database of \emph{WFPC2} PSFs, and a suitable 
match, HD 181204, was found in field position, colour, and exposure level. 
The sub-pixel registration, normalisation, and subtraction of 
the stellar PSF was performed following the method of \cite{Krist97}.
Figure~\ref{fig:obs_HST} shows the long F606W and F814W long exposures
after PSF 
subtraction. Obvious PSF artifacts remain in the subtracted 
image in the form of the saturation column bleed and residual 
diffraction spikes.

\subsubsection{HST/NICMOS observations and data processing}

Complementary near-IR observations were carried out on 2005 July 11
with the Near Infrared Camera and Multi-Object Spectrometer (\emph{NICMOS})
as part of the {\it HST} Cycle 13 GO/10177 program (PI: Schneider).
High contrast images of IM Lupi's circumstellar disc were obtained
using \emph{NICMOS} camera 2's coronagraphic imaging mode (spatial scale 75.8
mas.pixel$^{-1}$) following a strategy nearly identical to that used
to image the debris discs around HD 32297 \citep{Schneider05},  HD
181327 \citep{Schneider06}
but at a wavelength of 1.6 microns (F160W
filter $\lambda_\mathrm{eff}$ = 1.604 $\mu$m, FWHM = 0.401 $\mu$m,
spatial resolution of 0.16\arcsec).  

To mitigate the potentially degrading effects of localised, but
rotationally invariant, optical artifacts that appear in \emph{NICMOS}
coronagraphic images, IM Lupi was observed at two celestial orientation
angles offset by 29.86\degr.  IM Lupi was autonomously acquired and
optimally positioned behind the 0.3\arcsec\ radius coronagraphic
obscuration following 0.241\,s target acquisition imaging (also in the
F160W filter) at each of the two spacecraft (and field)
orientations. Following their respective target acquisitions, two sets
of F160W coronagraphic observations (each yielding 704\,s of
integration time after median combination of count-rate images derived
from three STEP32 multiaccum exposures) were obtained in a single
spacecraft orbit to minimise instrumentally induced temporal
variability in image structure on multi-orbit timescales.  Direct
(non-coronagraphic) images of IM Lupi (10.15\,s), were also obtained
after slewing the target out of the coronagraphic obscuration. 

A set of similarly observed, high SNR, coronagraphic
images of bright, isolated, non-disc bearing stars\footnote {As
  determined from earlier (HST Cycle 7) \emph{NICMOS} coronagraphic
  observations, specifically from GO/7226 for the PSF template stars
  discussed here.} observed in GO/10177, served as point spread
function (PSF) templates to create PSF subtracted images of the IM~Lupi
disc. This process reduced the stellar light below the enhanced contrast levels
provided by the coronagraph alone.  The PSFs of two of the template
stars with J-H and H-K colours similar to IM Lupi, GJ 3653 (M0.5V)
and GL 517 (K5V),
were well matched in structure to IM Lupi stellar PSF
impressed upon the disc images.  The deeply exposed F160W images of
the PSF templates (total integration times of 1352 s and 1224s,
respectively, for GJ 3653 and GJ 517) were flux density renormalised
to that of IM Lupi (based on target acquisition and direct imaging of all
three stars) and used to construct four PSF subtracted images of the
IM Lupi circumstellar disc (two from each orientation using both PSF
templates). Residual optical artifacts arising from the {\it HST}
secondary mirror support structure were masked in the individual
PSF-subtracted images.  All four images were then median combined
after re-orientation to a common celestial frame to produce the then
photometrically calibrated  image shown in Fig.~\ref{fig:obs_HST}
(right panel) and used in the
analysis discussed in this paper.  For further details of the technique
and process we refer the reader to the aforementioned disc imaging
papers: \cite{Schneider05} and \cite{Schneider06}.

\subsubsection{Disc Morphology, Brightness \& Scattering Fraction}

PSF subtracted images reveal, at all wavelengths, a 
compact nebula adjacent to the SW of the star. 
Morphologically, the nebula is a broad, gentle, symmetrical arc nearly
4\arcsec\ in size (celestial PA = 143\degr $\pm$ 5\degr). 
No clear differences are found between the nebulosity at 
0.606 and 0.814\,$\mu$m despite the presence of H$\alpha$ 
and other emission lines in the F606W filter. In the more sensitive
1.6 micron image, the ``nebulosity'' is (like in the \emph{WFPC2} images)
brightest to the SW, but  is clearly seen circumscribing the star.  The
scattered light to the NE is not apparent in the PC2 images partly due
to the instrumental sensitivity limitations in PC2 PSF subtraction
compared to \emph{NICMOS} coronagraphy with PSF subtraction. 

The IM Lupi images bear a striking
morphological resemblance to the protoplanetary disc around the T
Tauri star GM Aur \citep{Schneider03} and indicate the presence of a
circumstellar disc inclined in the range 40-60$^\circ$ towards the line of sight.  
Because of the similar appearance of the nebula in F606W, F814W and
 F160W, with a strong front/back asymmetry,  we 
conclude that it is dominated by scattered light from the star. 
To the SW of the elliptically shaped circumstellar nebulosity, an
arc-like dark band (presumably the higher opacity disc midplane
centered at r$\approx$1.5\arcsec along the disc morphological minor axis),
bifurcates an isophotally concentric lower surface brightness
scattering region (presumably the "lower" scattering surface of the
back side of the disc) extending to r$\approx$2.5\arcsec on the morphological minor axis of the
disc. This feature is most 
obvious in the F814W and F160W images. 
The dark lane between the upper and lower scattered light nebulae is
best seen
$\sim1.4$\arcsec\ southwest of the star. This coincides with the
probable forward scattering direction given the purported disc
inclination (see Fig.~\ref{fig:schema}).

The total brightness of IM~Lupi in the saturated \emph{WFPC2} images was
measured using
the method of \cite{Gilliland94}.  The results are 0.088 $\pm$ 0.0044\,Jy
(F606W$_\mathrm{mag}$=11.44 $\pm$ 0.05) and 0.150 $\pm$ 0.0075\,Jy
(F814W$_\mathrm{mag}$=10.52 $\pm$ 0.05)  in the F606W and F814W filters respectively.
The fraction of light scattered by the disc (relative to the
  stellar flux at this wavelength) is 0.0186 in the F814W
image, measured in the region beyond 0.2\arcsec radius from the star.

\begin{figure}
  \includegraphics[width=\hsize]{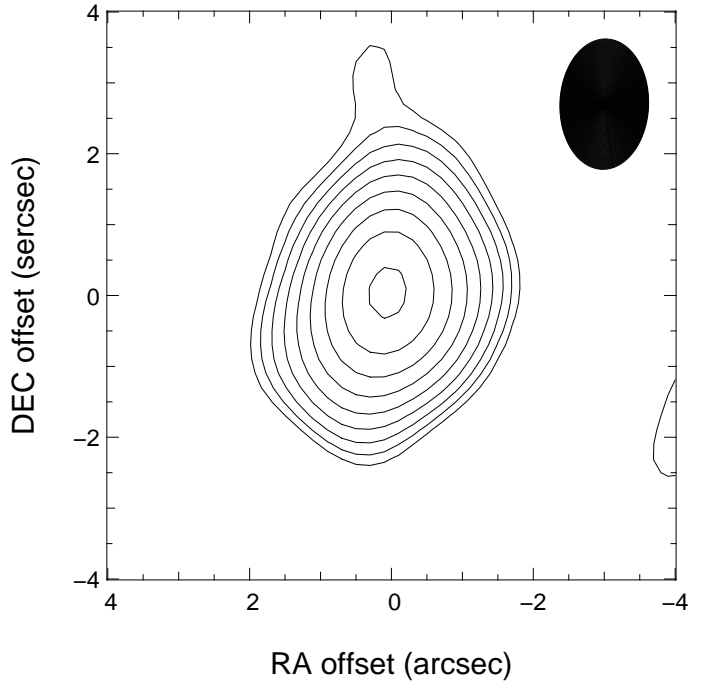}
  \caption{\emph{SMA} 1.30\,mm aperture synthesis image of IM~Lupi. Contours
    begin at 
 6\,mJy/beam and step in factors of $\sqrt{2}$ in intensity.
 The black
    ellipse shows the CLEAN half-intensity beam size. The orientation
    is the same as in Fig.\,\ref{fig:obs_HST}. The zero position is RA
    = 15:56:09.203, DEC = -37:56:06.40 (j2000).
    \label{fig:sma_map}}
\end{figure}

\begin{figure}
  \centering
  \includegraphics[width=\hsize]{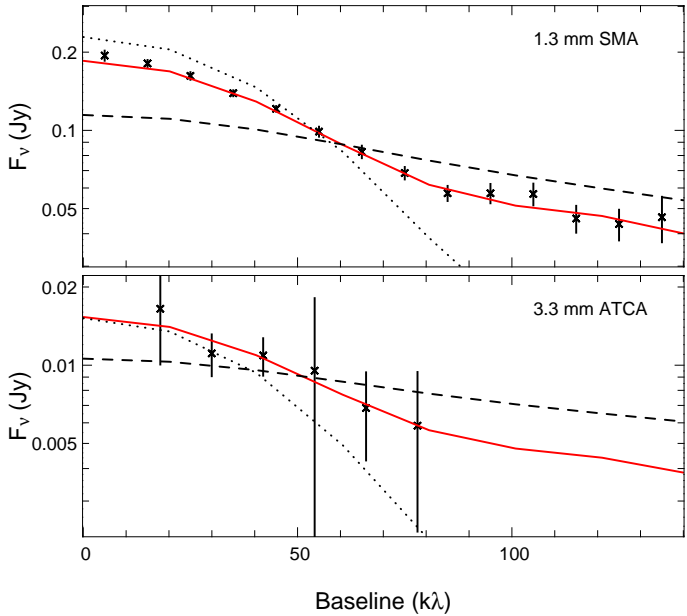}
  \caption{Circularly averaged 1.3\,mm (upper panel) and 3.3\,mm (lower panel) visibility data (crosses) compared
    to the models (lines) with different surface density
    exponents. The best fit model is shown in the full red line and
    corresponds to $\alpha=-1$. The dashed and dotted lines correspond to the best
    models with $\alpha=-2$ and 0, where the inclination is enforced
    to be $50^\circ$ (see section~\ref{sec:best_model}).
   \label{fig:Vis_mm}}
\end{figure}

From the PSF-subtracted coronagraphic image, the
measured F160W-band disc flux density at 0.3\arcsec\ $> r >$ 4.5\arcsec\
is 11.8\,mJy (uncertainty $\approx$ 4\,\%).
This flux density excludes the small near-central area, shown in black
in Fig.~\ref{fig:obs_HST}, which is
insufficiently sampled due to masked optical artifacts.
  From the direct and target acquisition
images we measure the F160W-band stellar flux density as 0.616 $\pm$
0.012\,Jy (F160W$_\mathrm{mag}$=8.11 $\pm$ 0.02) via
TinyTim\footnote{http://www.stsci.edu/software/tinytim/tinytim.html}
\citep{Krist97b} model PSF fitting and aperture photometry.
Thus, the fraction of 1.6 $\mu$m starlight scattered by the disc
(beyond 0.3\arcsec\ from the star) is 0.019 $\pm$ 0.001, consistent
with the F814W value.

In the F160W image, a large scale fainter halo is sensitively detected to a
distance of $\approx$ 4\farcs4 from the central star, along the
morphological major axis (see Fig.~\ref{fig:schema}). This large scale nebula may indicate the presence of a tentative
envelope surrounding the circumstellar disc. Because the dark lane
and counter nebula of the disc, which are seen through the potential
envelope, are detected, this envelope must be optically thin at
optical wavelengths. It is
not detected in the F606W and F814W images. In the following, we focus
on the disc properties and do not try to reproduce the halo in our modelling. 

\subsection{Spitzer IRS and MIPSSED spectra}

  IM Lupi was observed with the \emph{IRS} spectrograph installed
  onboard the \emph{Spitzer Space Telescope} as part of the ``Cores to
  Discs'' (\emph{c2d}) legacy program
  (AOR: 0005644800, PI:~Evans, \citealp{Kessler-Silacci06}). 
  The observations took place on 
  2004 August 30 using the four proposed modules (Short-Low,
  Long-Low, Short-High and Long-High), corresponding to a wavelength coverage
  of 5.2-38.0$\,\mu$m with a spectral resolution R between 60-127 for
  the two ``low'' modules and R $\sim$ 600 for the two ``high'' modules. 
 The data reduction was performed using the \emph{c2d} legacy team pipeline
 \citep{Lahuis2006} with the S13 pre-reduced (BCD) data, using the aperture
 extraction method\footnote{The extraction was done following two different methods: 
full aperture extraction and PSF extraction. PSF extraction is 
less sensitive to bad data samples but for some modules the 
estimated PSF is subpixel-sized. As a consequence, the PSF extraction
becomes unstable. In this paper,we adopt the spectrum 
obtained with the full aperture extraction method because of its 
better stability.}.
  Pointing errors of the telescope can produce offsets
 between the different modules which were corrected for by the
 reduction pipeline. 
Table~\ref{tab:IRS_obs} summarises the details of the observations
 while the spectrum is presented in Fig.~\ref{fig:silicates}.

 \begin{table}
\caption{Details of the \emph{Spitzer/IRS} observations. SL, SH, LL,
  and LH refer to Short-Low, Short-High, Long-Low, Long-High
  respectively. R is the spectral resolution and SNR the signal to
  noise ratio. 
  \label{tab:IRS_obs}} 
\begin{tabular}{c|c|c|c|c}
\hline
\hline
 Module & R & Integration time & SNR & Pointing\\
   &  & (t$_\mathrm{int}\times$n$_\mathrm{dce}\times$n$_\mathrm{exp}$) & &  offset (\arcsec)\\
\hline
 SL & 60-127 & 14 $\times$ 1 $\times$ 2 & 20 & -0.8 -- 0.2 \\
 SH & $\sim$ 600 & 31 $\times$ 2 $\times$ 2 & 71 & 0.7 -- 1.3 \\
 LL & 60-127 & 14 $\times$ 1 $\times$ 2 & 35 & -1.4 -- -0.3 \\
 LH & $\sim$ 600 & 60 $\times$ 1 $\times$ 2 & 66 & 3.0 -- 1.2 \\ \hline
\end{tabular}
 \end{table}

The $52$--$97\,\mu$m MIPSSED data ($R \sim 15$--$25$) were taken on
2006 March 31 (Program ID 1098, PI: Kessler-Silacci). The basic
calibrated data (BCDs) were coadded using the 
MOsaicking and Point source EXtractor (MOPEX) software. Because the
flux from IM Lupi is weak, the automated MOPEX extraction failed to
extract the flux. MOPEX indeed focused on the edge of the detector,
where there are a lot of hot pixels, rather than on the signal from the
source. The MIPSSED spectrum was therefore extracted with IRAF since
it allows the user to set the centre of the aperture (around columns
12-16) and performs an optimised extraction. The resulting spectrum remains
too noisy to detect any spectral features, such as the
70\,$\mu$m cystalline emission band.
In order to increase the signal-to-noise and get a better estimate of the
mid-IR slope of the SED, we decided to bin the spectrum to calculate 3 photometric measurements
at $\sim$ 60, 75 and 90\,$\mu$m (see Table~\ref{tab:photometry}).

\begin{figure}
 \includegraphics[width=\hsize]{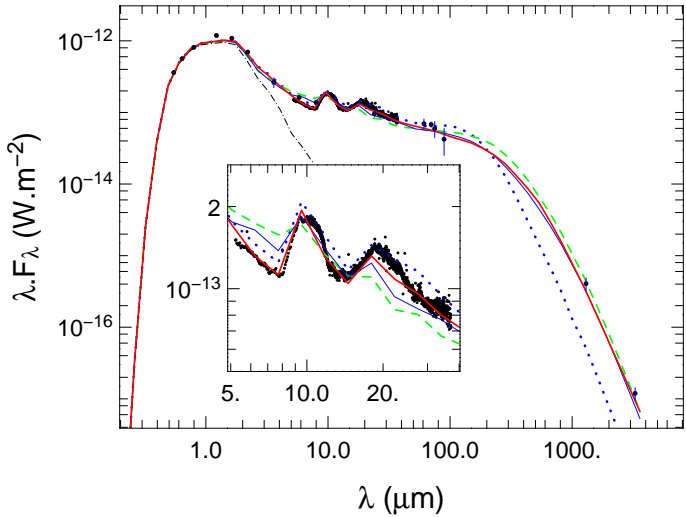}%
 \caption{Qualitative analysis of the SED of IM~Lupi.
    The blue dotted line presents a SED calculated with submicron particles ($a_\mathrm{max} =
    3\,\mu$m). Silicates emission bands are well reproduced but
    the millimetre fluxes are too low, even with a disc mass of
    $M_\mathrm{disc}=0.1\,M_\odot$.
    A model with millimetre grains ($a_\mathrm{max} =
    3\,$mm, green dashed line) performs much better at long wavelengths
    but the silicate features then disappear. A model
    including millimetre grains ($a_\mathrm{max} =
    3\,$mm) close to the midplane and submicron grains in the upper
    layers (\emph{i.e.} with a stratified structure as described in section~\ref{sec:differenciation}), can reproduce both
    the silicates bands and the millimetre
    fluxes (red full line). The thin black dot-dash line represents
    the stellar photosphere.\label{fig:IMLup_qualitatif}}
\end{figure}

The IRS spectrum of IM Lupi clearly shows silicate emission features typical
of Class II objects \citep{Hanner98, Kessler-Silacci06}. In addition to the amorphous features at 10\,$\mu$m 
(Si--O stretching mode) and 18\,$\mu$m (O--Si--O bending mode), other 
features are visible in the spectrum that we attribute to 
crystalline forsterite (Mg--rich end member of the olivine group) and 
crystalline enstatite (Mg--rich pyroxene, \citealp{Koike00,Molster02}). 
The 9.3\,$\mu$m feature, attributed to
enstatite, is clearly visible, as is the crystalline complex at around 
27\,$\mu$m. This latter complex is likely a blend of a forsterite feature at 27.6\,$\mu$m plus
an enstatite feature at 28.2\,$\mu$m (Figure~\ref{fig:silicates}). A
complex around 24$\,\mu$m may also be tentatively detected at a
signal-to-noise between 1 and 3. This complex is possibly a blend of 
enstatite features at 23.8 and 24.5$\,\mu$m and of a forsterite feature
at 24$\,\mu$m. The forsterite feature at 33.6$\,\mu$m is not observed. 

The shape and strength of the amorphous 10\,$\mu$m feature
are related to the mean size of the emitting grains. These have been used
as observational diagnostics for grain growth in discs (e.g.
\citealp{vanBoekel05} and  \citealp{Kessler-Silacci06}).
We measured the shape and strength of the IM\,Lupi 10\,$\mu$m feature
following the approach described in \cite{Kessler-Silacci06}.
The continuum normalisation is obtained using their first method.
The fluxes at the reference wavelengths, $S_{11.3}$ and $S_{9.8}$, are
then calculated by integrating over a wavelength range of
0.1\,$\mu$m centered on $\lambda=9.8\,\mu$m and 11.3\,$\mu$m. Finally,
the
feature strength is estimated by calculating the mean peak flux,
$S_\mathrm{Peak}$, of the normalised spectrum. 
We obtain a $S_{11.3}/S_{9.8}$ ratio of 1.00 and a $S_\mathrm{Peak}$ of 1.60,
which locates IM\,Lupi in the bulk of the class II objects observed by 
 \cite{Kessler-Silacci06} with a flat 10\,$\mu$m feature, consistent
with the presence of micron-sized grains. 
Overall, these features indicate a significant level of processing of the
silicate grains in the disc compared to those observed in the
interstellar medium, which instead are found to be submicron-sized and largely amorphous ($<$1\,\% of
crystalline silicates, \citealp{Kemper04}).

\subsection{Spatially resolved SMA observations of 1.3\,mm continuum
  emission}
\label{sec:SMA}

The 1.3\,mm (230.538\,GHz) thermal dust emission from IM~Lupi was
observed with the Submillimeter Array\footnote{The Submillimeter Array
  is a joint project between the Smithsonian Astrophysical Observatory
  and the Academia Sinica Institute of Astronomy and 
Astrophysics and is funded by the Smithsonian Institution and the
Academia Sinica.} (\emph{SMA}). See Pani\'c et al (in prep) for the
observational details and full analysis. The extended
configuration of the \emph{SMA}, with baselines up to 180\,m, provided
an excellent sampling of the \emph{uv}-plane. The corresponding
beam is 1.84\arcsec $\times$ 1.25\arcsec, with a position angle of 0.2$^\circ$.

Figure~\ref{fig:sma_map} presents the
CLEAN map of IM Lupi's disc. The disc is clearly resolved with FWHMs
of 2.33 $\times$ 1.72\arcsec\ and a position angle of 167$^\circ$.
 Taking into account the beam elongation, this PA
 is in agreement with the PA of 143$^\circ$ derived from
scattered light images. The emission extends to $\approx$ 2\arcsec
from the star at the 3 sigma level of 6.28\,mJy/beam.

The total integrated 1.3\,mm flux is 175.8$\pm$4\,mJy. 
Together with the 3.3\,mm flux of 13\,mJy measured by
\cite{Lommen07}, this leads to an estimate of the millimetre spectral index
$\alpha_\mathrm{mm} = \dd\log(F_\nu)/\dd\log(\nu) = 2.80 \pm 0.25$.

Because of the relatively large noise of individual visibilities in the
\emph{uv}-plane, a direct model fitting to these is not very
informative, and some form of averaging is necessary. 
Figure~\ref{fig:Vis_mm} shows the amplitude
averaged over \emph{uv}-distance in concentric circular annuli. 
%(blue) and
%elliptical annuli (red). 
We compare this amplitude with the amplitude averaged in elliptical
annuli of eccentricity $e = 0.65$
(corresponding to a axisymmetric structure observed at an
inclination of 50$^\circ$). 
The latter is aimed at providing a more correct
representation of the data, taking into account projection effects
\citep{Lay97}. The difference in amplitude for the two
averaging methods is smaller than the rms levels of the observations
at all baselines. Thus, the deprojection of the visibilities is not
necessary in this study. In the following, we adopt the amplitude averaged in
concentric circular annuli and use the same procedure to calculate the
visibilities of the different models. 

\subsection{Spectral energy distribution}

IM~Lupi is an irregular variable \citep{Batalha98} and has been observed to flare 
dramatically in  U~band \citep{Gahm93}.
To limit the effect of
variability in our analysis, we adopt the optical photometry 
presented by \cite{Padgett06}, which is contemporaneous to the mid-IR
measurements.
Mid- and far-infrared flux densities for IM~Lupi were measured by the
\emph{c2d} legacy project \citep{Evans03}.  
A detailed description of the reduction and source
extraction procedures used for the \emph{IRAC} and \emph{MIPS}
measurements can be found in \cite{Harvey06,Harvey07} and in the
\emph{c2d} delivery document \citep{Evans07}. 
All measurements are quoted in Table~\ref{tab:photometry}.

\begin{table}
  \caption{Photometric measurements of IM Lupi. References: (1)~=~\cite{Padgett06}, (2) = \cite{Lommen07}\label{tab:photometry}}
  \centering
  \begin{tabular}{ccccc}
    \hline
    \hline
    $\lambda$ ($\mu$m) & Flux (Jy) & Date & Ref.\\
    \hline
    0.545&  0.0655 $\pm$ 0.0007 & 06/2004 & (1) \\
    0.638&  0.120  $\pm$  0.0012 & 06/2004& (1) \\
    0.797&  0.216  $\pm$  0.0022 & 06/2004 & (1) \\
    1.22 & 0.483   $\pm$  0.0048 & 03/06/2000 & 2MASS \\
    1.63 & 0.591   $\pm$   0.0059 & 03/06/2000 & 2MASS \\
  2.2 & 0.511   $\pm$  0.0051 & 03/06/2000 & 2MASS \\
   3.6   &    0.324 $\pm$  0.0184 & 12/08/2004&  Spitzer/IRAC   \\
   4.5   &    0.220 $\pm$  0.0178 & 12/08/2004&  Spitzer/IRAC  \\
   5.8   &    0.313 $\pm$  0.0156 & 12/08/2004&   Spitzer/IRAC\\
   8.0   &    0.370 $\pm$  0.0223  & 12/08/2004&   Spitzer/IRAC\\
  24   &    0.765 $\pm$  0.0708 &  02/08/2004&  Spitzer/MIPS\\
 61.1    & 1.42   $\pm$ 0.22  &  31/03/2006&  Spitzer/MIPSSED\\
  70     &    1.581 $\pm$  0.127& 02/08/2004&   Spitzer/MIPS\\
 74.8    & 1.48    $\pm$ 0.37 & 02/08/2004&   Spitzer/MIPSSED\\
 89.3    &   1.26  $\pm$ 0.51 & 02/08/2004&   Spitzer/MIPSSED\\
1300  &    0.1758$\pm$  0.0351 & 21/05/2006 &   SMA     \\
3276  &    0.013 $\pm$  0.0026 & 08/2005 & (2) \\ 
  \hline
  \end{tabular}
\end{table}

%===================================================================================
\section{Simple estimation of some of the star and disc parameters\label{sec:direct_estimation}}
\label{sec:simple_estimation}

Each of the previously presented observations gives, by itself, strong
insights into the properties of 
IM~Lupi's disc and its dust content. Our goals in the following
sections are to exploit this complete data set in order to draw complementary
constraints and obtain a finer understanding of the circumstellar
environment of IM~Lupi and  its evolutionary state.
We aim at building a as
coherent a picture as possible, by analysing the different observations
simultaneously, in 
the framework of a single model. 

Even with simple assumptions to
describe the disc structure and dust properties, exploring the
whole parameter space is far beyond current modelling capabilities. 
Instead, the analysis was performed in 2 steps: 
\begin{enumerate}
\item We extract, whenever possible, parameters ``directly'' from
  observations or via simple modelling: disc
  size from scattered light images, disc dust mass and maximum grain
  size from the millimetre spectral index and dust composition from
  mid-infrared spectroscopy (this section). 
  We also fit the SED alone, by a manual exploration of the parameter space, to test the
  need for a spatial differentiation of dust grains within the
  disc. 
  The goal of this preliminary analysis is to determine which parameters can be
  kept fixed, allowing us to significantly reduce the dimensionality of
  parameter space to be explored, and which parameters need to be varied
  in our modelling.
\item We then systematically explore these remaining parameters, which
  cannot be easily extracted from observation and/or 
  can be correlated with each other. 
  We calculate a full grid of models and perform a simultaneous fit to all
  observations (section~\ref{sec:grille}).
\end{enumerate}

\subsection{Model description}

Synthetic images and spectral energy distributions  are computed using
MCFOST, a 3D continuum radiative transfer code based on the 
Monte Carlo method \citep{Pinte06}. MCFOST includes multiple 
scattering with a complete treatment  of polarization (using the 
Stokes formalism), passive dust heating assuming radiative equilibrium,
and continuum thermal re-emission. 
In short, the code first computes the temperature structure assuming
that the
dust is in radiative equilibrium with the local radiation field. This
is done by propagating
photon packets, originally emitted by the central star, through a
combination of scattering  
(following Mie theory), absorption and reemission events until they exit the
computation grid. This step uses the algorithms described in
\cite{Bjorkman01} and \cite{Lucy99} with a total number of 1\,280\,000
photon packets.  
The SEDs are then computed by emitting and propagating the proper amount of
stellar and disc thermal photon packets, ensuring that at least
12\,800 packets are contributing in each wavelength bin. 
All of these packets are allowed to scatter in the disc as often as
needed. We use 102 wavelength bins, globally distributed in a
logarithmic scale between 0.3 and 3\,500\,$\mu$m but with an increased
resolution (linear scale with bins of 0.5\,$\mu$m) between 5 and
40\,$\mu$m, to properly sample the regime observed with the \emph{IRS} spectrograph. 
About 4 millions packets were used to compute each of the scattered light
images and thermal emission maps
The maximum wavelength of \emph{HST} observations being 1.6\,$\mu$m, we assume that the disc
thermal emission is negligible in the scattered light images. For the millimetre maps, both the
disc thermal emission and star contribution are considered.

We consider an axisymmetric flared density structure with a Gaussian vertical
profile $\rho(r,z) = \rho_0(r)\,\exp(-z^2/2\,h(r)^2)$ valid for a
vertically isothermal, hydrostatic, non self-gravitating disc. We use
power-law distributions for the surface density $\Sigma(r) =
\Sigma_0\,(r/r_0)^{\alpha}$  and the scale height $ h(r) = h_0\,
(r/r_0)^{\beta}$ where $r$ is the radial coordinate in the equatorial
plane and $h_0$ is the scale height at the radius $r_0 =100$ AU.
The disc extends from an inner cylindrical radius  $r_\mathrm{in}$  to an
outer limit $r_\mathrm{out}$.

Dust grains are defined as homogeneous and spherical 
particles (Mie theory) with sizes distributed 
according to the power-law $\dd n(a) \propto a^p\,\dd a$, with $a_{\mathrm{min}}$
and  $a_{\mathrm{max}}$ the 
minimum and maximum sizes of grains. Extinction and scattering opacities, scattering
phase functions, and Mueller matrices are calculated using Mie theory.
We adopt a
power-law index $ p = -3.5$, which is 
%\TODO{3.5 ou 3.7 ??? : 3.7 sur simu1 et 3.5 sur simu2} 
generally used to reproduce interstellar
extinction curves. We choose a minimum grain size
$a_\mathrm{min}$, small enough that its exact value has no effect
(we  fix it to 0.03\,$\mu$m). The maximum grain size $a_\mathrm{max}$
is considered as a free parameter.

\subsection{Star properties} \label{sec:star_prop}

\cite{Krautter92} discussed the 
available distance estimates for the Lupus star forming regions, ranging from 130 up to 
300\,pc, and concluded that the most likely distance is between the 
limits 130 and 170\,pc.  The most recent determinations are from 
\cite{Hughes93}, who proposed a distance of 140$\pm$20\,pc  and from
 \cite{Wichmann98}, who concluded a distance of 190$\pm$27\,pc from {\it
   Hipparcos} parallax. In the following, we adopt this value
 of 190\,pc and discuss the effect of distance uncertainties on the parameters
 derived from our modelling in section~\ref{sec:validity_range}.

IM~Lupi has been classified as an M0 star from photometric measurements
by \cite{Hughes94}. 
The estimated age for this system ranges from $10^5$\,yr to $10^7$\,yr
\citep{Hughes94}, depending on the pre-main sequence evolutionary 
track used. 
In the following modelling sections, we adopt a NextGen stellar
atmosphere model \citep{Allard97} with 
3\,900\,K and $\log(g) = 3.5$, a stellar radius of 3\,R$_\odot$ (L = 1.9\,L$_\odot$) and
an $A_\mathrm{V}$ of 0.5\,mag, which matches the optical photometry of IM Lupi.
Following the star evolution models of \cite{Baraffe98}, this roughly
corresponds to a 1\,M$_\odot$ star with an age of $10^6$\,yr.

{\it Hipparcos} observations suggested a companion with a separation of only 
0\farcs4 \citep{Wichmann98}. However, neither near-IR speckle
observations \citep{Ghez97} nor the current {\it HST} observations
confirm the presence of a close companion.  The extended nebulosity
may have been marginally detected by {\it Hipparcos} and incorrectly
interpreted in terms of a binary.

\subsection{Dust composition}

The strength and shape of the $10$ and $20\,\mu$m emission silicate
features give us some information on the dust grain composition. We
find that micron sized amorphous
olivine \citep[Mg\,Fe\,SiO$_4$, refractive indices from][]{Dorschner95} dust grains
reproduce in a qualitative way 
the emission bands. In the following general modelling approach, 
we do not try to match perfectly  the shape of the features, but instead their strength, and we fix
the  composition of dust to $100\,\%$ amorphous olivine. 
A detailed analysis of the silicates features is presented in
section~\ref{sec:mineralogie}. The results of this analysis are
consistent with the composition we adopted here.

\subsection{Disc outer radius}
  
Scattered light images reveal the presence of a disc with a diameter of
approximately $\approx $ 4\arcsec. The fact that we see the dark lane and
the corresponding counter-nebula, produced by scattered light from the
backside of the disc, indicates that the disc, if it
remains optically thick at optical wavelengths, cannot extend much further away than 400\,AU
(assuming a distance of 190\,pc). In the following we fix the outer
radius in our models to $400$\,AU.  

\subsection{Disc mass}

Basically, the millimetre fluxes give access to the product of the disc
dust mass with the dust absorption opacity.
Assuming amorphous olivine spherical grains \citep{Dorschner95},
we obtain a maximum absorption opacity at 1.3\,mm of
2\,cm$^2$ per gram of dust, corresponding to a grain size distribution with a
maximum grain size close to 3\,mm. This value is also the value
adopted by \cite{Beckwith90} and more recently  \cite{Andrews07}.
With this opacity, the millimetre visibilities are
reproduced by a dust disc mass of
$10^{-3}$\,M$_\odot$. This mass is estimated using the
temperature structure from the solution of the radiative
equilibrium.
Grain size distributions with a significantly
smaller or larger maximum grain size have a
lower opacity at millimetre wavelengths and would imply larger disc dust
masses to account for the observed fluxes.
With the assumption that the gas to dust ratio is 100, as in molecular
clouds, the minimum total disc mass is
$0.1$~M$_\odot$. 
With a star mass close to 1\,M$_\odot$, the
disc mass cannot be much larger than $0.1$~M$_\odot$. We
then consider it also as a maximum value and fix the disc mass to this
value. 
It should be noted that millimetre dust opacities remain
poorly known. The disc mass should thus be understood as an
estimate with an uncertainty of a factor of a few.

\subsection{Assessment of the need for grain size stratification\label{sec:need_for_settling}}
\label{sec:differenciation}

The SED of IM~Lupi presents some clear features that we can use to
guide our analysis on the dust properties.
Our modelling confirms that the disc is almost entirely optically thin in the millimetre
regime. We can therefore use the millimetre spectral index of $2.80
\pm 0.25$ to constrain the
grain size and, thus the corresponding millimetre opacity slope is
$\beta_\mathrm{mm} = 0.8 \pm 0.25$. 
This value is
significantly lower than the expected value between 1.5 and 2 for
interstellar medium (ISM) grains ($\leq 1\,\mu$m), which 
 indicates that larger grains, of the order of 1\,mm in size, are
present in the disc.
However, the IRS spectrum clearly shows
strong silicate emission bands that cannot be reproduced by a simple
power-law grain size distribution extending to millimetre sizes, even
if it contains a high fraction of micron-sized grains. 
The silicate emission requires that the maximum grain size do not exceed a few microns
 in warm (close to the star), optically thin regions of the disc at
mid-IR wavelengths (surface).
In this section, we explore whether a dust population that is
perfectly mixed with the gas, and hence uniform in the disc, can
simultaneously reproduce both of these features, or whether a stratified structure with
spatially separated dust populations of small and large grains is
required to solve this apparent contradiction.

Assuming that the dust and gas
are perfectly mixed, the silicate emission features are best
reproduced with a maximum grain size $a_\mathrm{max} \leq 3\,\mu$m,
essentially because larger grains are featureless at mid-IR wavelengths.  
Fig.~\ref{fig:IMLup_qualitatif} (blue dotted line) presents such a synthetic SED,
with $a_\mathrm{max} = 3\,\mu$m. However, this model ($\beta_\mathrm{mm} = 4$) fails to
reproduce the observed millimetre spectral index, as
expected for grains in the Rayleigh regime.
Furthermore, because of the
low opacity of micron size grains in the millimetre regime, we also find a 
millimetre flux too low, by a factor of 10 relative to observations.

The  spectral index between $1.3$ and
$3.3$\,mm suggests instead the presence of at least millimetre-sized
grains. Thus, a grain size distribution with a maximum grain size
close to  $a_\mathrm{max} = 3$\,mm (Fig.~\ref{fig:IMLup_qualitatif},
dashed green line) is in good agreement 
with our millimetre observations. This kind of grain size distribution however, does not
reproduce the silicate emission 
features which almost completely disappear. 
Indeed, for a grain size distribution with a slope $p=-3.5$ and a
  maximum grain size $\gtrsim 100\,\mu$m,  the mid-IR
opacities are not dominated by the sub-micron sized grains but rather by the
grains a few microns in size which have featureless opacities
(see for instance Fig.~5 of \citealp{Kessler-Silacci06}).
We can then conclude that
millimetre grains are present in the disc, but that they do not dominate
the opacity in regions producing silicate 
emission features: inner edge and disc surface at radii
smaller than a few astronomical units.  
A spatial dependence of the dust properties, particularly
grain size distribution, seems to be required to account for all
observables. In the
following, we assume that this spatial dependence is produced by vertical dust
settling in the disc.  This structure is characterised by small grains
on the disc surface, which produce the silicate emission, and larger
grains in the midplane which emit most of the thermal radiation in the
millimetre regime.  We discuss
other possible explanations for the grain size stratification in section \ref{sec:discussion}. We
describe this dust stratification  by a simple parametric law, the reference
scale height being a function of the grain size
\begin{equation}
  h_\circ(a) = h_\circ(a_\mathrm{min}) \,
  \left(\frac{a}{a_\mathrm{min}} \right)^{-\xi}\ .
\end{equation}
In the case where the dust and gas are perfectly mixed, $\xi=0$, and
$h_\circ(a) = h_\circ(a_\mathrm{min})$ is independent of the grain
size and is equal to
the scale height of the gas disc. 

The full red line in Fig.~\ref{fig:IMLup_qualitatif} presents a
SED calculated with $a_\mathrm{max} = 3$\,mm and the stratification
described above, taking $\xi = 0.1$. This kind of model reproduces
 both the silicate emission bands and
millimetre fluxes and gives support to the presence of a stratified
structure in the disc of IM~Lupi. 
We would like to emphasise that, at this point, this model is not a fit, just
an educated guess as far as disc parameters are concerned, meant only
to illustrate the impact of settling on the SED.  A similar effect was
also found by  \D04 and \cite{D'Alessio06}.
In the following, we fix the maximum grain size to $a_\mathrm{max} =
3$\,mm to match the spectral index and consider $\xi$ as a free parameter.

%======================================================================
\begin{figure}
  \includegraphics[width=\hsize]{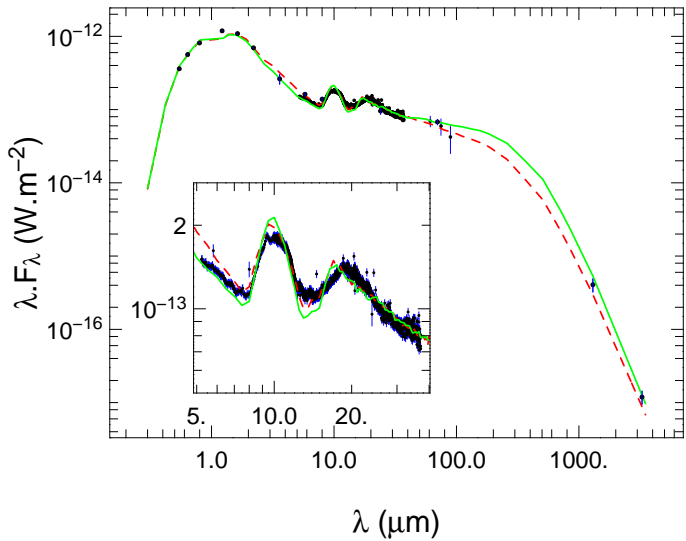}
  \caption{SEDs of the best models. The dashed red line corresponds to
    best model for the fit of the SED only and the full green line
    corresponds to the simultaneous fit of all observations. \label{fig:best_SED}}
\end{figure}

\section{Inferred disc model parameters from multi-technique model fitting\label{sec:grille}}
\label{sec:grid_modelling}

Encouraged by the previously described qualitative behaviour, we then
moved to a quantitative analysis by exploring a full grid of models. 
The goal is to find a single model which self-consistently fits all
observables and estimates the 
parameters as well as their range of validity and uncertainties. 
Because the F606W and F814W images are very similar, we restrict our
fitting of the scattered light images to the F606W and F160W
bands. Since they are the shortest and longest wavelengths: the
intermediate F814W image is likely to be well fitted by a model that
adequately represents these 2 wavelengths. 

\subsection{Explored parameter space}

Exploration of the parameter space was performed by varying the
geometrical properties of the disc ($r_\mathrm{in}$, $\beta$,
$\alpha$, $h_\circ(a_\mathrm{min})$) and the degree of the dust
settling $\xi$.  Scattered
light images give us a good idea of the inclination angle, around
50$^\circ$, but we consider it as a free parameter in order to fine tune our estimation. 
This results in a total of 6 independently varied parameters.
The range explored for each parameter is summarised in
Table~\ref{tab:param}. All combinations of parameters were explored,
resulting in a total of 421\,200 calculated models.
The ranges were chosen to sample physically plausible
values for the different parameters, and adapted during the course of
the modelling to calculate all the models which may be a reasonably good
representation of data. The probability curves we obtain (Figure
\ref{fig:proba_all_IMLup}) for the 
different parameters show that we have sampled a large enough fraction of the
parameter space. 

\begin{table}
  \caption{Ranges of parameters explored in our grid of models\label{tab:param}}
  \centering
  \begin{tabular}{lllll}
   \hline
   \hline
   parameter & Min. & Max  & N$_\mathrm{sampl}$ & Sampling \\
   \hline
   $i$ ($^\circ$)& 0 &  90 &  10 & linear in $\cos$ \\ 
   $\alpha$ & -2.0 & 0.0 & 5 & linear\\
   $\beta$ & 1.0 & 1.25  & 6 & linear\\ 
   $r_\mathrm{in}$ (AU) & 0.05 & 1.5 & 12 & logarithmic\\
   $h_\circ(a_\mathrm{min})$ (AU)& 8 & 16 & 13 & linear\\
   $\xi$ &  0.0 & 0.2 & 9 & linear\\
   \hline
  \end{tabular}
\end{table}

\begin{figure*}[thbp]
  \includegraphics[height=0.28\hsize]{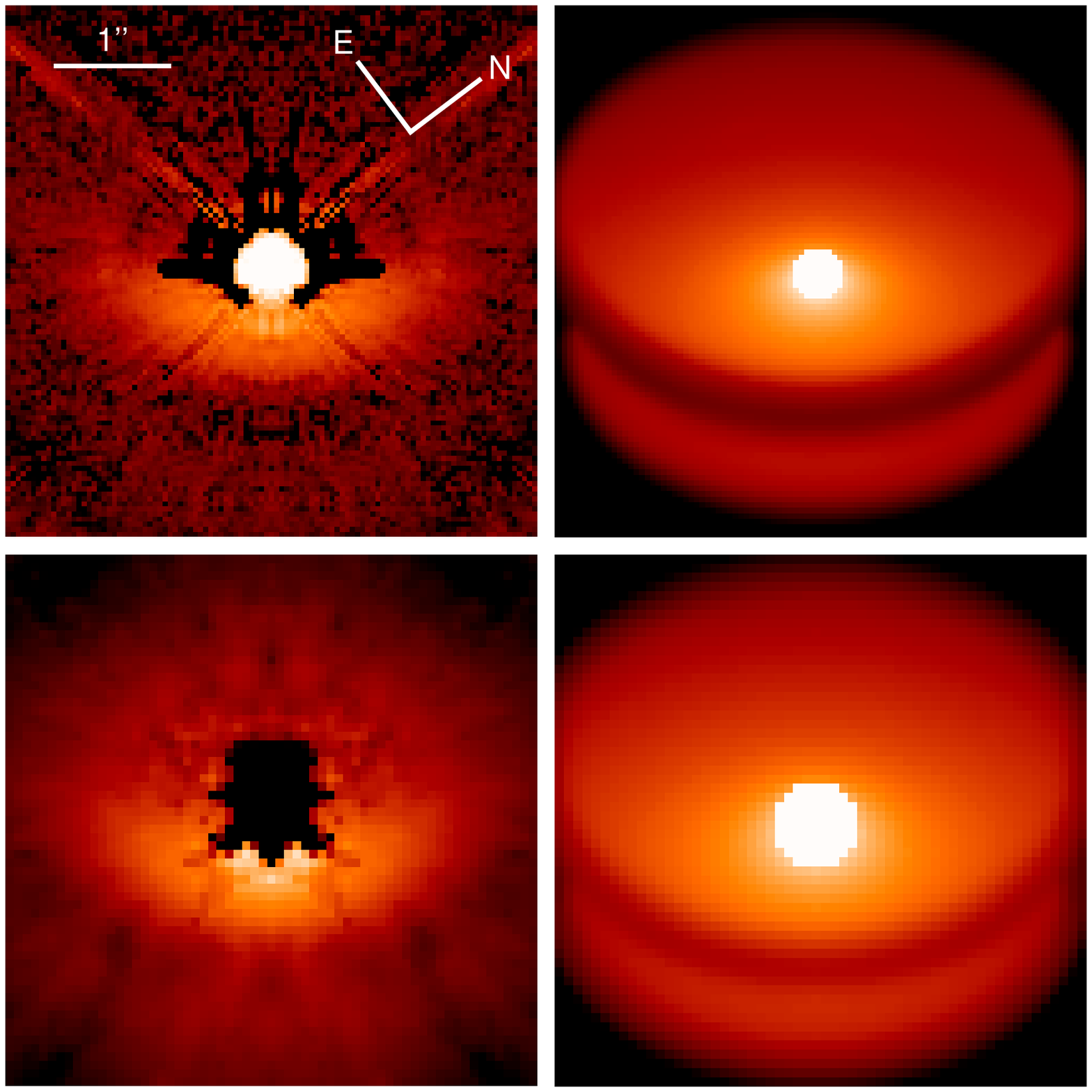}
  \hspace{\stretch{1}}
  \includegraphics[height=0.27\hsize]{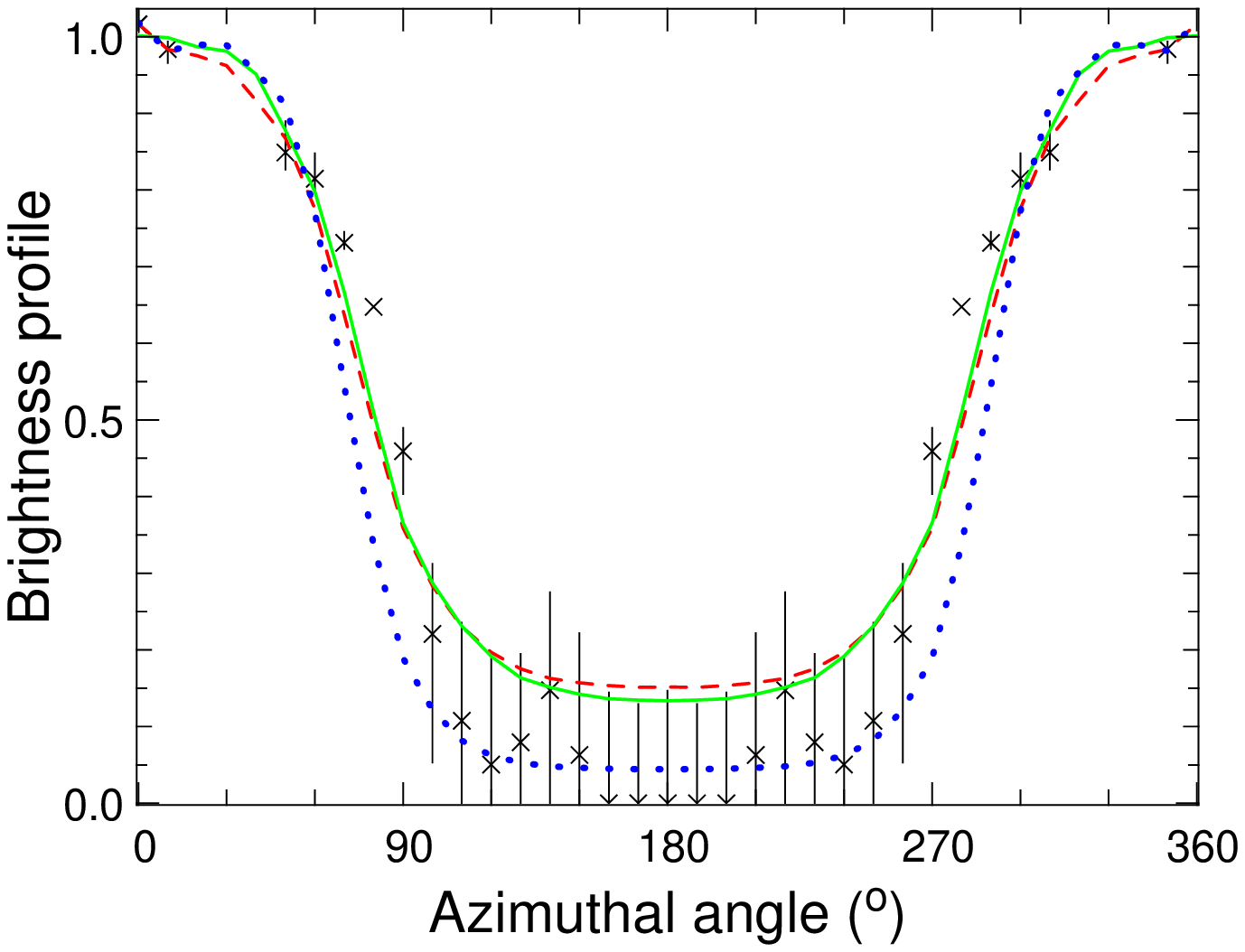}
   \includegraphics[height=0.27\hsize]{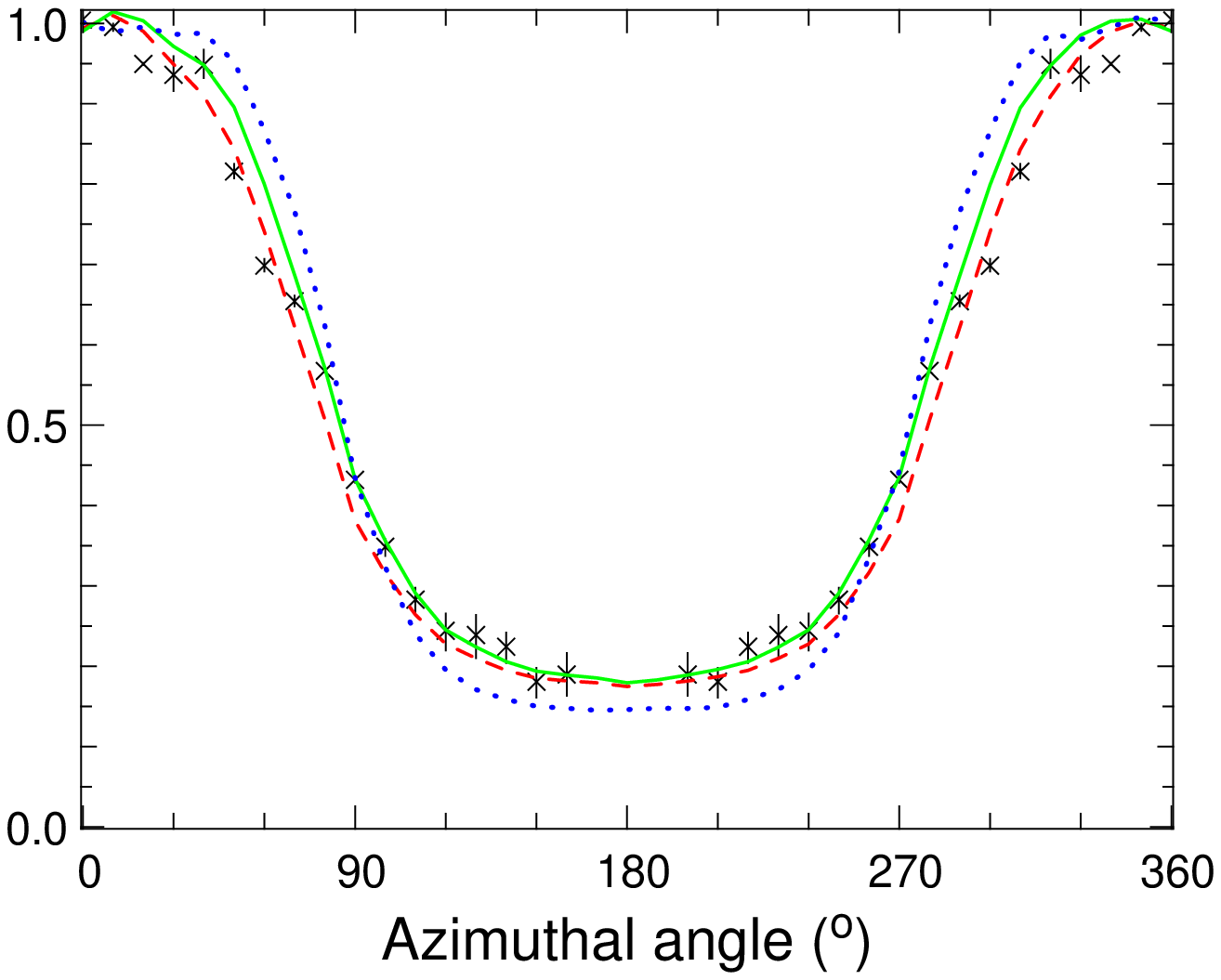}

\caption{Scattered light images of the best models compared to
    observations. {\bf Left panel:} The upper row corresponds to the images at
    0.606\,$\mu$m and the lower row to those at 1.6\,$\mu$m. Synthetic
    maps (right) were convolved by the core of the PSF (up
    to 5 pixels from the peak). {\bf Central panel:} 0.6\,$\mu$m azimuthal brightness
  profile.  {\bf Right panel:} 1.6\,$\mu$m azimuthal brightness
  profile. In both central and right panels, the red dashed line
  correspond to the best fit of the scattered light only, the full
  green line to the best fit of all observations simultaneously and
  the blue dotted to the scattered light images with porous
  grains. The azimuthal angle is 0$^\circ$ in the front side of the
  disc, \emph{i.e.} towards bottom in the first panel. 
\label{fig:best_IMLup_scatt}}
\end{figure*}

\subsection{Fitting procedure}

The fitting of the SED is performed through the definition of a
reduced $\chi^2$: $\chi^2_\mathrm{SED}$ from the observed
and synthetic fluxes, and from the observational uncertainties. Because
synthetic SEDs are sampled with a large number of packets,
we neglect the model Monte Carlo error bars which are significantly
smaller ($< 1\,\%$). We can get a sense of their amplitude in the
silicate feature (inset in Fig.~\ref{fig:best_SED}).
 
Comparing synthetic images with observations is a more delicate
procedure due to the
presence of the star that dominates the flux in the synthetic images.
  Observational effects
are taken into account by choosing pixel sizes similar to those of
observations, \emph{i.e.} 0.045\,\arcsec/pixel at 0.6\,$\mu$m and 0.076\,\arcsec/pixel at 1.6\,$\mu$m. Real scattered light images were
obtained by manual subtraction of the PSF.  Such a procedure cannot
be considered for the models however, 
given the large number of models dealt with here.
To make the problem of PSF subtraction tractable, we choose to convolve
our models only with the PSF core (up to 5 pixels from the peak) in order to reproduce its
smoothing effect, but at the same time avoiding the superimposition of
the convolved
direct stellar light on the disc, where it is detected in the actual
PSF-subtracted images.
The disc appears left-right symmetrical in both {\it WFPC2} and {\it NICMOS} images. To
increase the signal-to-noise ratio and to reduce artifacts, the observed images were
symmetrised relative to the semi-minor axis of the disc, prior to
comparison with models. 

A pixel-by-pixel comparison between models and observations is very
sensitive to observation artifacts. It requires a precise map of the
observational uncertainties in the image in order to make optimal use
of its information. This procedure is complex 
for high contrast PSF subtracted images. Large deviations in small regions
of the image can strongly bias the fitting procedure. Also, PSF
subtraction uncertainties introduce systematic, as opposed to random,
errors, which are not properly taken into account in a $\chi^2$
minimisation. 
Instead, we adopt  an image fitting  
performed by extracting geometrical observables. Our goal here is to
reproduce the main characteristics of the images, allowing
small parts of the model images to not be in perfect agreement with the
observations, if necessary.
We therefore adopt the following method, illustrated in 
Fig.~\ref{fig:best_IMLup_scatt}, which is based 
on the azimuthal intensity variations.
For both wavelengths, we calculate the observed intensity variations 
normalised to the average flux in the azimuthal angle range
[-15$^\circ$,+15$^\circ$] (centered on the disc's semi-minor axis). 
For each 10$^\circ$ sector, we extract the average flux of the pixels
encompassed between two ellipses of eccentricity $e = 0.65$
(corresponding to a circularly-symmetric structure observed at an
inclination of 50$^\circ$) and of 
semi-major axis of 1.5 and 2.1\arcsec.
Regions presenting diffraction artifacts are moreover excluded from these sectors.
Uncertainties were calculated by repeating the same procedure on the
observations from which we have added and subtracted the estimated
error maps.
These error maps are constructed by taking the shot noise
from photon statistics and adding in a PSF scaled by the uncertainty factor in the
normalisation used for PSF subtraction. 

The synthetic brightness profiles were extracted in the same way from
Monte Carlo images. We do not consider any uncertainties for these
synthetic profiles.
For the
1.6\,$\mu$m image we further use
the positions of the dark lane and second nebula as geometrical
observables to fit for. 
We compare the synthetic maps and the observations using  reduced
$\chi^2$ based on the previously defined observables\footnote{A total
  of number of 31 and 34 measurements were used at 0.6 and 1.6\,$\mu$m respectively.}:
$\chi^2_{0.6\,\mu\mathrm{m}}$  and $\chi^2_{1.6\,\mu\mathrm{m}}$.  
 
Model comparison with \emph{SMA} and \emph{ATCA} data was performed in
$(u,v)$ plane to avoid additional uncertainties from the image
reconstruction with the CLEAN algorithm. Synthetic visibilities were
computed by Fourier transform of the MCFOST emission maps. 
The visibilities were then binned in circular annuli as done for the
observed visibilities, and the reduced $\chi^2$ is computed from
these binned visibilities: $\chi^2_\mathrm{mm}$. 
Because only 6 data points are available for the \emph{ATCA}
observations, it is not possible to define a meaningful reduced  $\chi^2$ for
these observations alone. Instead, we fit simultaneously the \emph{ATCA} and
\emph{SMA} data by defining a unique  reduced  $\chi^2$.
To avoid an eventual contamination by a potential envelope and/or surrounding
cloud, fitting was only performed for baselines larger than 40\,k$\lambda$
which probe spatial scales smaller than 5\,\arcsec, \emph{i.e.}
corresponding to the disc observed in scattered light.

We compute a total $\chi^2_\mathrm{tot}$ defined as the sum of all
the reduced $\chi^2$:
\begin{equation}
  \label{eq:chi2}
  \chi^2_\mathrm{tot} = \chi^2_\mathrm{SED} +
  \chi^2_{0.6\,\mu\mathrm{m}} + \chi^2_{1.6\,\mu\mathrm{m}} + \chi^2_\mathrm{mm}.
\end{equation}
 This $\chi^2_\mathrm{tot}$ allows us to do a global
fitting of all observations simultaneously.  

%======================================================================
\subsection{Best model}
\label{sec:best_model}
The parameters of the model with the lowest $\chi^2_\mathrm{tot}$ are
described in Table~\ref{tab:best_parameters}. Overall, the best fit model is in very
good agreement with all observations, especially taking into account
the range of wavelengths and the variety of observations analysed in
the fitting procedure.  Table~\ref{tab:chi2} shows the $\chi^2$ corresponding to
best models fitting only one of the observations and to the global
best model. The global best models has $\chi^2$ values only slightly
larger than the $\chi^2$ of the best models fitting each of the
observations, showing that it offers a good representation of all observations.

The SED of this model is represented in Fig.~\ref{fig:best_SED}.
It is in very good agreement with observations from the optical to
the millimetre regimes. In particular, it reproduces
both the silicate bands and millimetre fluxes, and hence the
millimetre spectral index. 
The silicates features
are roughly reproduced but the shape is not exactly identical to the
observed one, resulting in relatively high values of the $\chi^2$. 
This is due to our very simple approximation on optical
properties, with dust grains only composed of amorphous silicates. We
recall here that a precise match of the silicate emission bands was not our goal at
this stage of the modelling and that we were only interested in the
amplitude of the silicate features.

Synthetic images of the best model are compared with observations on  
Fig.~\ref{fig:best_IMLup_scatt}. The general shape is well reproduced
at both wavelengths, with the right roundness, brightness
distribution and a dark lane in good agreement with observations.
The central and right panels show a quantitative comparison of the
azimuthal brightness profiles. The agreement is good as shown by
the values of the $\chi^2$ in Table~\ref{tab:chi2}. The model predicts
profiles that are very similar at both wavelengths whereas the
observations suggest (given the relatively low SNR and large PSF
uncertainty) a stronger contrast at 0.6\,$\mu$m, with a back 
side compatible with a non-detection. We discuss this point in detail in
section~\ref{sec:porosite}.

Figure~\ref{fig:Vis_mm} shows the results of the fit of the 1.3 and
3.3\,mm visibilities. The agreement is also very good for the
baselines $> 40\,$k$\lambda$ on which the fitting was performed. At
smaller baselines, the best models are marginally below the data
points. This may be due to a small contamination of the observations by the surrounding
molecular cloud and/or envelope.
Millimetre visibilities are mainly sensitive to the apparent spatial
distribution of the dust, i.e, 
the surface density exponent $\alpha$, the disc outer radius, the
total mass and the disc inclination. 
With an outer radius of  400\,AU derived from scattered
light images and a dust disc mass based on the millimetre flux, our
modelling allow us to constrain the surface density profile (Fig.~\ref{fig:Vis_mm}). 
The $\alpha$=-1 model (full line) provides an excellent match. Conversely,
the other two models ($\alpha$=0 and $\alpha$=-2) fail to
successfully reproduce the data, if we force the inclination to be
close to 50$^\circ$. Models with  $\alpha$=0 provide a good match to
the observations for a disc closer to edge-on (inclination of $\approx
70^\circ$), which is incompatible with the scattered light images.

\begin{table}[htbp]
 \caption{Best model parameters. The best model is the model with the
    lowest $\chi^2_\mathrm{tot}$. The valid range for each parameter
    is defined as the
    range of values which symmetrically enclose the central 68\,\% of
    the probability (see text for details). This is equivalent to a
    $1\,\sigma$ confidence interval.\label{tab:best_parameters}}
  \centering
  \begin{tabular}{lll}
    \hline
    \hline
    parameter & best model & valid range\\
    \hline
    $T_\mathrm{eff}$ (K)& 3\,900& fixed\\
    $R_\mathrm{star}$ ($R_\odot$) & 3.0& fixed\\
    distance (pc) & 190 & fixed\\
    $r_\mathrm{out}$ (AU)& 400 & fixed\\
    $M_\mathrm{dust}$ ($M_\odot$)& $10^{-3}$ & fixed\\
    dust composition & \multicolumn{2}{l}{100\,\% amorphous olivine, fixed}\\
    $a_\mathrm{min}$ & 0.03\,$\mu$m & fixed\\
    $a_\mathrm{max}$ & 3\,mm & fixed\\
    $i$ ($^\circ$)& 50 & 45 -- 53\\
    surface dens. exp &  -1 & -0.84 -- -1.21\\
    flaring exponent & 1.15 & 1.13 -- 1.17\\
    $r_\mathrm{in}$ (AU)& 0.32\,AU  & 0.25 -- 0.4\\
    $h_\circ(a_\mathrm{min})$ & 10\,AU & 9.5 -- 10.3\\
    settling exponent $\xi$ & 0.05 & 0.02 -- 0.07\\
    \hline
  \end{tabular}
 \end{table}

In Fig.~\ref{fig:best_SED} and \ref{fig:best_IMLup_scatt}, we present both
the overall best fit model and best fits to single observables. 
This quantitatively shows that the best overall fit is very similar to
the individual fits and illustrates how limited single data sets are: families
of models can be found that give equally good fits to the
observations. It is only through a combination of the various observations 
that we can solve most ambiguities and 
disentangle between models (Fig.~\ref{fig:proba_all_IMLup}).

\subsection{Validity range of  parameters}
\label{sec:validity_range}

\begin{figure*}[tbp]
  \psfrag{cos(i)}[c][c]{\Large $\cos (i)$}
  \psfrag{surface density}[c][c]{\Large surface density $\alpha$}
  \psfrag{flaring}[c][c]{\Large flaring $\beta$}
  \psfrag{rin}[c][c]{\Large $r_\mathrm{in}$ (AU)}
  \psfrag{H}[c][c]{\Large $h_0$ (AU)}
  \psfrag{stratification}[c][c]{\Large settling $\xi$}
  \psfrag{Probability}[c][c]{\Large Probability}
  \includegraphics[width=\hsize]{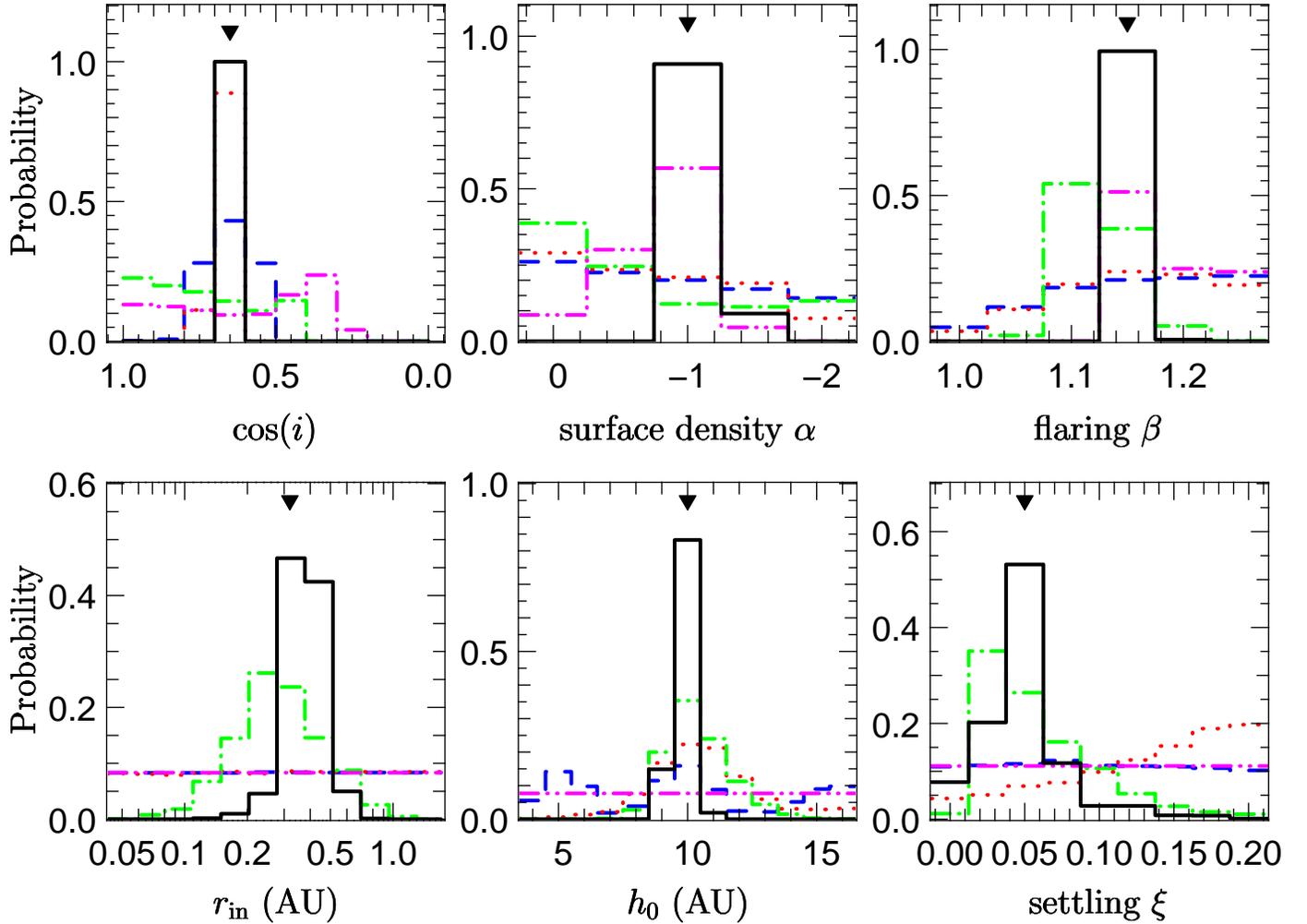}
  \caption{Bayesian probabilities of the various parameters for the
    scattered light images at 0.6\,$\mu$m (dashed blue) and
    1.6\,$\mu$m (dotted red),
    the SED (dot-dash green), the millimetre visibilities
    (dot-dot-dash pink) and for the
    images, SED and mm visibilities simultaneously (full black
    line). The triangles represent the parameters of the best model.
\label{fig:proba_all_IMLup}}
\end{figure*}

To determine the range of validity of the different parameters, we use
a Bayesian inference method \citep{Press92,Lay97,Pinte07}.
 This technique allows us to estimate the probability
of occurrence of each value of a given parameter.  
 The relative probability of a point of the parameter space
 (\emph{i.e.} one model)
is given by $\exp (-\chi^2/2)$ where $\chi^2$ refers to the previously
defined reduced
$\chi^2$ of the corresponding model. All probabilities are
normalised at the end of the procedure so that the sum of the
probabilities of all models over the entire grid is equal to 1.
This method analyses all the models in a statistical sense and does
not give any specific attention to a given model, including the best model.

The Bayesian method relies on \emph{a priori}
probabilities on the parameters. Here, we assume that we do not have any
\emph{preliminary available information}, and we choose uniform a priori
probabilities which correspond to a uniform sampling of
the parameters. However, in the absence of any data, some grid points are
more likely than others: consideration on solid angles show that an
inclination between 80 and 90$^\circ$ (close to edge-on) is more
likely than a inclination between 0 and 10$^\circ$. Uniformly
  distributed disc inclinations and orientations in three dimensions 
correspond to a uniform distribution in the cosine of the inclination.
Also, some physical quantities,
like the inner radius, tend to be distributed logarithmically. The
grid was built according to these distributions (Table~\ref{tab:param}).

Figure~\ref{fig:proba_all_IMLup} presents the relative figure of
merit estimated from the Bayesian inference for each of the parameters,
 for the fitting of the scattered light
images, SED, and millimetre visibilities. These results were obtained
by marginalising (\emph{i.e.} summing) the
probabilities of all models where one parameter is fixed successively to its different
values, \emph{i.e.} the probabilities of our 6-dimension parameter
space are projected successively on each of the
dimensions (they are not cut through the parameter space).
Potential correlations
between parameters are entirely accounted for with this approach, and
the error bars extracted from the probability curves take into
account the interplay between parameters. 
The resulting histograms represent the
probability that a parameter takes a certain value, given the data and
assumptions of our modelling. 

Both scattered light images are similar and naturally lead to similar
probability densities (blue dashed and red dotted lines). Because the
error bars are smaller for the \emph{NICMOS} image, the resulting
constraints on the parameters are stronger. 
The disc
inclination is well constrained in 
the range  45 -- 55$^\circ$. Fitting of both images tends to prefer
flaring exponents $\geq 1.1$. 
 Because the disc is
optically thick at optical and near-infrared wavelengths, scattered
light images are insensitive to the surface density.
These images probe the outer part  of the disc, and due to the
flared geometry adopted in the current work, the stellar illumination 
does not depend on the inner radius.
As a consequence, we do not get any
constraint on the inner edge of the disc.
The azimuthal brightness profile is strongly sensitive to the disc
scale height. Both images give slightly different, but compatible,
constraints on the scale height, with a peak of probability at 10\,AU.
The probability curve for the 0.6$\,\mu$m image also presents secondary
peaks at 5 and 15\,AU. These peaks correspond to the two inclination
bins adjacent to the most probable bin centered at $\cos(i) = 0.65$.  

The analysis of the SED gives different constraints (green dot-dash line in
Fig.~\ref{fig:proba_all_IMLup}). As expected, the
inclination is not well constrained. Only the models with high
inclination, \emph{i.e.} models that occult the star, are excluded.
As the disc is very massive, it remains opaque to its own radiation up
to long wavelengths, and the SED is only slightly sensitive to the disc surface
density index.
The flaring index, describing the disc geometry and its
capacity to intercept stellar light, is strongly constrained with most
probable values between 1.1 and 1.15. The inner radius is constrained
between 0.15 and 0.6\,AU. The probability curve for the
settling index is of particular interest. The value 0, corresponding
to the case without settling, has a significantly lower probability than those
corresponding to a relatively low amount of settling. The peak of
probability is around $\xi =0.025$ and is followed by a decreasing probability density
towards strongly settled discs. The probability becomes negligible for
$\xi > 0.15$. Indeed, high degrees of settling result in 
silicate features that are too strong and  mid-infrared fluxes that
are too low.
The scale height is constrained to be around 10\,AU, in agreement
with the modelling of the scattered light images.

The pink dot-dot-dash line shows the constraints resulting from the
fitting of the millimetre visibilities. As expected, because they are
mainly sensitive to the bulk of the disc mass, these
observations do not give any constraints on the inner radius, scale
height or degree of dust settling. 
Because our fitting was realised
on azimuthally averaged visibilities, the inclination is not well
constrained. Only the close to edge-on disc models are excluded. As
previously mentioned however, the CLEAN map does
give an inclination consistent with the one deduced from scattered
light images. The main constraints are obtained on i) the flaring index,
with the most probable value between 1.1 and 1.2, which is in agreement  with the
results of the SED modelling and ii) on the surface density index,
with a peak of probably close to $-1$, and the extreme values of 0 and $-2$
being excluded. There is a strong correlation between the surface
density index and the inclination, models with $\alpha > -1$
corresponding to inclinations $> 55^\circ$. Indeed, the millimetre
interferometers probe the apparent spatial distribution of the dust
and an inclined disc can mimic a disc with a more concentrated dust
distribution. 

Combined, the various observations give complementary
constraints on the model parameters.
Figure~\ref{fig:proba_all_IMLup} illustrates, in a quantitative way,
the complementarity between
the results extracted from 
the scattered light images, the SED, and the millimetre
visibilities.
 The Bayesian analysis gives the relative likelihood
$p_i(M|D_i)$ of different models $M$, given the data $D_i$ (scattered light
images, SED or millimetre visibilities) that have been
measured.  Given the $n$ different data sets, the relative likelihood
of a model is then:
\begin{equation} 
p(M|D_1 \ldots D_n) = \prod_{i=1}^n\, p_i(M|D_i),  
\end{equation}
because the uncertainties in each data set are independent. In our
case, this is equivalent to calculate the probability from 
a $\chi^2_\mathrm{tot}$ we have defined as the sum of the individual reduced
$\chi^2$.
The black line in Fig.~\ref{fig:proba_all_IMLup} shows the resulting
marginalised probabilities, corresponding to the
simultaneous fitting of the two scattered light images, the SED, and
the millimetre visibilities. All parameters are constrained  within narrow ranges and the corresponding
probability curves are much sharper, \emph{i.e.} the constraints on
the parameters are stronger (see Table~\ref{tab:best_parameters}). This illustrates the need for
simultaneous modelling of various kinds of observation to
quantitatively derive disc parameters and obtain finer models. 
Table~\ref{tab:best_parameters} gives the range of validity of the
different parameters. For each parameter $\theta$ with a density of
probability $p(\theta)$, this range of validity is defined as 
the interval [$\theta_1$, $\theta_2$] where
\begin{equation}
  \int_{\theta_\mathrm{min}}^{\theta_1} p(\theta)\,\dd \theta =
  \int_{\theta_2}^{\theta_\mathrm{max}} p(\theta)\,\dd \theta =
  \frac{1-\gamma}{2}  
\end{equation}
with $\gamma = 0.68$. The interval [$\theta_1$, $\theta_2$] is a
68\,\% confidence interval and corresponds to the $1\,\sigma$ interval
for a Gaussian density of probability.

The triangles in Fig.~\ref{fig:proba_all_IMLup} represent  the
different parameters of the best model. Although this is not necessarily
the case in Bayesian studies, it coincides here with the most
probable model, defined as the model with the parameters which have
the highest probabilities. This means that the best model
corresponds to the global minimum of $\chi^2$ in the parameter space. 
There is no contradiction between the constraints from the different observations
indicating that our model, although based on simple assumptions,
provides an adequate description (at the precision level of present
observations) of the disc structure and dust
properties.

\begin{table}
  \caption{Reduced $\chi^2$ values for the best models.\label{tab:chi2}}
  \begin{tabular}{llllll}
    \hline
    \hline
    Model & $\chi^2_{0.6\,\mu\mathrm{m}}$ &
    $\chi^2_{1.6\,\mu\mathrm{m}}$ &  $\chi^2_\mathrm{SED}$ &
    $\chi^2_\mathrm{mm}$ & $\chi^2_\mathrm{tot}$\\
    \hline
      Best $0.6\,\mu$m &    {\bf 1.16} &      17.24 &     83.47 &     83.60 &    185.47 \\ 
      Best $1.6\,\mu$m &     3.02 &       {\bf 0.48} &     29.75 &     18.64 &     51.89 \\ 
      Best SED &         22.83 &     139.53 &      {\bf 9.87} &     51.24 &    223.48 \\ 
      Best mm &      22.69 &      46.13 &    560.75 &      {\bf 0.43} &    630.00 \\ 
      Best all obs.  &     {\bf 1.56} &       {\bf 0.68} &     {\bf
        14.02} &      {\bf 1.18} &     {\bf 17.44} \\ 
    \hline
  \end{tabular}
\end{table}

\subsection{Effect of distance}

The results described thus far have been determined using a distance
of 190\,pc. It should however, be noted that the distance of IM Lupi
is relatively uncertain (see section \ref{sec:star_prop}). An alternate distance
value will thus affect the results of our modelling. Fortunately, due
to the self-similarity of the radiative transfer equations, our
results can be scaled for an alternate distance. If instead the
distance to IM Lupi is \mbox{$A\times190$\,pc}, where $A$ is a real
positive number, the calculations will be mathematically identical to
the ones that we have presented  if (i) all lengths
in our model (star radius, disc inner and outer radius $r_\mathrm{in}$
and $r_\mathrm{out}$) are multiplied by $A$, (ii) the geometry is
not modified (the scale height at 100\,AU, flaring and surface density
indices, degree of dust settling remain identical) and (iii) the disc mass
is multiplied by $A^2$ (ensuring that the disc opacity is not
modified).

\section{Detailed analysis of the dust properties}
\label{sec:dust_properties}
In the previous section, we performed a global fit to all of the
observations simultaneously. Such a fit however cannot account for the
fine signatures we detect in the various observations. In this
section, we analyse in detail the silicate emission bands and
scattered light images to infer additional information on the dust
properties.  

\subsection{Disc mineralogy}
\label{sec:mineralogie}

\subsubsection{Compositional fitting of the 10\,$\mu$m silicate feature}

A detailed study of the 10\,$\mu$m feature provides quantitative information on the
composition and size of the grains responsible for the observed
emission (e.g. \citealp{Bouwman01}, \citealp{vanBoekel05}, 
\citealp{Apai05}, \citealp{Schegerer06} or more recently \citealp{Merin07},
\citealp{Bouy07} and \citealp{Bouwman08}).
% showed that we are able to probe 
%dust composition in certain regions of the disc.
To fit the 10\,$\mu$m feature, we use five dust species:
olivine (glassy  Mg\,Fe\,SiO$_4$, \citealp{Dorschner95}), pyroxene
(glassy  Mg\,Fe\,SiO$_6$, \citealp{Dorschner95}) and 
silica (amorphous quartz (silicon dioxide) at 10K,
\citealp{Henning97}) for the amorphous grains, and enstatite
\citep{Jaeger98}  plus forsterite \citep{Servoin73} for 
the crystalline species.
 Furthermore, we consider two single representative
grain sizes (generally 0.1 and 1.5\,$\mu$m, although other grain sizes have
been tested) following previous studies. Once the
continuum emission has been subtracted, a fit to the amorphous 10\,$\mu$m feature is calculated. This is done
by multiplying grain opacities with relative mass fractions and with a
single-temperature blackbody. A pseudo-$\chi^{2}$ minimisation procedure
provides the best relative mass fractions for all of the species and
grain sizes, as well as the best temperature.

Achieving a good estimation of the dust continuum emission contributing
to the spectrum is the most difficult part for this sort of analysis, because
it influences the derived composition. 
We learned from previous studies that a simple
local continuum of the 10\,$\mu$m feature is not a good solution, despite
the fact that it is the simplest one. We find that it is necessary to leave flux 
at the end of the amorphous feature, at around 13.5\,$\mu$m, because amorphous silicate 
grains are still contributing to the emission at these wavelengths.
We therefore adopt the continuum produced by the MCFOST code to obtain a more
realistic estimate of the continuum shape around
10\,$\mu$m. Nevertheless, the best-fit disc model includes amorphous silicate grain 
emission, and consequently the SED is not featureless. We generated another
continuum by artificially removing the amorphous features in the MCFOST
calculations. Refractive indices in the wavelength range 7-35\,$\mu$m, which
includes the silicates features, were replaced by a $\log-\log$
interpolation with wavelength of the refractive indices from the values at $\lambda=$7 and 35\,$\mu$m. 
This solution, although providing a more physical continuum, remains
imperfect. First, the interpolated refractive indices are very
likely to be a strong simplification of what would be the refractive
indices without silicate features. Second, the energy normally escaping through the 
amorphous features is redistributed over adjacent wavelengths and this 
overestimates the flux near the feet of the amorphous features. A
smoothing treatment is thus required to obtain a good estimate of
the continuum.

\subsubsection{Results}

We investigated the influence of the grain size on the goodness of the
fit, the result of which is that only grains of an approximate size of 1.5 $\mu$m are
able to reproduce the amorphous feature (inset in
Fig.~\ref{fig:silicates}).
 A first fit with grain sizes
of 0.1 and 1.5\,$\mu$m reveals that 94\,\% of the grains need to have a
size of 1.5 $\mu$m. A second run assuming grain sizes equal to 1.5
and 3.0\,$\mu$m indicates that 97\,\% of the grains have a size of
1.5\,$\mu$m. Because 3.0\,$\mu$m grains are not featureless at
10\,$\mu$m, our results on the low fraction of these grains are robust.
This confirms the qualitative result we obtained
with the ratio $S_{11.3}/S_{9.8}$ versus the $S_\mathrm{Peak}$, which indicated
micron-sized grains responsible for the 10\,$\mu$m feature. The result is
very robust against using different continua. The derived temperature of these $1.5\,\mu$m grains is always around
350\,K. The best fit is obtained for a temperature of 357\,K.

In the radiative transfer modelling of the disc, the flux
received by the observer in the $10\,\mu$m silicate band mainly
comes  from grains slightly smaller than 
1$\,\mu$m in size~: 50\,\% of the energy is emitted by grains between 0.1 and
1.2$\,\mu$m and the the peak of the emission originates from grains
0.9$\,\mu$m in size. This explains the slightly sharper silicate
features in the model compared to observations
(Fig.~\ref{fig:best_SED}). 
The average temperature of the
grains responsible for the received emission at 10$\,\mu$m is
690\,K, which is significantly higher than the value for the fit of the
silicate feature. This may indicate that grains around
1.5\,$\mu$m, responsible of most of the emission in the silicate
band, are located at
larger radii and/or in deeper regions than what is assumed in the
global modelling. However, the large difference between
temperatures should be interpreted with care and should be
considered indicative only. Indeed, contrary to the global
modelling, the detailed composition fit of
the silicate feature has been performed after subtraction of the
continuum from the MCFOST calculations (then at a temperature of
690\,K). This continuum represents about 50\,\% of the flux at a wavelength of
10\,$\mu$m.

In order to reproduce the amorphous silicate feature, the fraction of
crystalline grains is always less than 10\,\%. For the best fit,
the crystallinity fraction is 7\,\%. %6.95\%. 
This can mainly be explained by
the presence of the enstatite feature at around 9.3\,$\mu$m. 
Long-wavelength features, due to enstatite and forsterite grains, also
indicate the presence of crystalline grains in cooler regions of the
disc (larger radii and/or deeper in the disc) than the regions probed
at 10\,$\mu$m \citep{Merin07}. However, these features
remains weak. No features are detected at wavelengths larger than
30\,$\mu$m which also suggests a low degree of crystallinity.

 \subsection{Are we observing fluffy aggregates?}
 \label{sec:porosite}
 Light scattering strongly depends on the dust properties, with the
 general trends that grains significantly smaller that the wavelength
 produce isotropic scattering and that the larger the grains the more
 anisotropic the scattering phase function becomes. All of IM~Lupi's scattered light
 images present a strong front-back asymmetry indicating anisotropic scattering.
 This implies the presence of grains at least comparable in size
 with the wavelength, in the regions of the disc probed by scattered
 light, \emph{i.e.} the disc surface at large radii ($\gtrsim$ a few 10
 AUs).
 Interestingly, the front-back flux ratio is higher at the shortest
 wavelength indicating that scattering could be more forward throwing in that
 case. Our modelling has lead to solutions that are marginally
 compatible with the flux level in the back side of the F606W image,
 although with a systematic overestimation (Fig.~\ref{fig:best_IMLup_scatt}).
 In this section, we explore what kind of dust grains could
produce a more contrasted azimuthal brightness profile at
0.6\,$\mu$m, whilst remaining in good agreement with the azimuthal brightness profile at
1.6\,$\mu$m, \emph{i.e.} we try to find dust properties that could
produce a
flux ratio $F(180^\circ) /  F(0^\circ)$ of 0.05 at 0.6\,$\mu$m and of
0.2 at 1.6\,$\mu$m (180$^\circ$ and 0$^\circ$ refer to the azimuthal
angles as in Fig.~\ref{fig:best_IMLup_scatt}, and not to the scattering
angles). 
As our best model is already marginally compatible
with the observations, we do not try to perform a new fit to all
observations simultaneously, instead only focusing on the effect of the dust
properties on the flux ratios. We adopt the geometry of the previously
estimated best model. We do not consider dust settling here and take
$a_\mathrm{max}$ as a free parameter. This value should now be
understood as the maximum grain size in the disc surface.

\begin{figure}
   \includegraphics[width=\hsize]{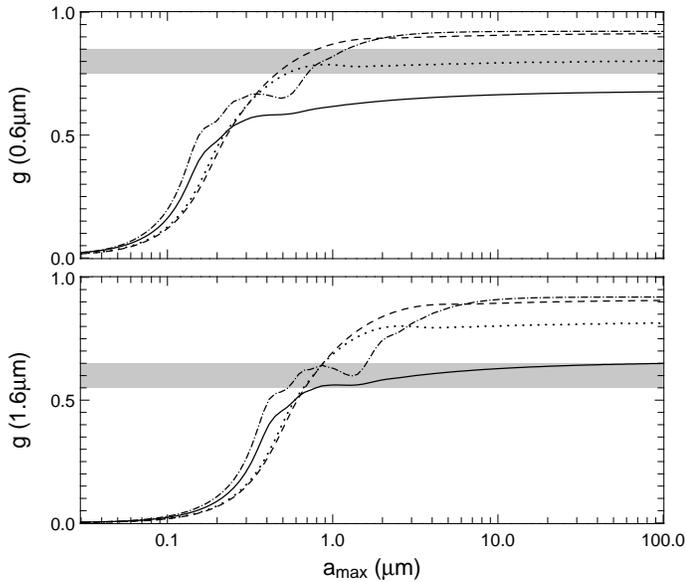}   
   \caption{Asymmetry parameter as a function of maximum grain size
     for different dust compositions. The top panel is for $\lambda =
     0.6\,\mu$m and the bottom panel for $\lambda = 1.6\,\mu$m.
     The full line corresponds to the olivine compact grains used in
     the global modelling (slope of the grain size distribution
       $p=-3.5$). The dashed line corresponds to the same grains,
       with a porosity of 80\,\% ($p= -3.5$). The dot-dashed line
       corresponds to compact grains with p = 0.
 The dotted line corresponds to pure water ice grains.
     The shaded areas represent
     the region that would fit the flux ratio of both the F606W and
     F160W scattered light images. A model reproducing both scattered
     light images must lie on the shaded area of both panels for the
     same maximum grain size.\label{fig:g_vs_g}}
 \end{figure}

 The scattering anisotropy can be described, as a first approximation,
 in terms of the 
 asymmetry parameter $g_\lambda = \left< \cos \theta_\lambda \right>$
 where $\theta_\lambda$ is the scattering angle at the wavelength
 $\lambda$. The models we have presented so far correspond to $g$
 values around 0.6 for both wavelengths, whereas for a better
 agreement on the flux on the backside of the disc, models with 
 $g_{0.6\mu\mathrm{m}} \approx 0.8$ are needed. In
 Fig.~\ref{fig:g_vs_g}, we plot the region ($g_{0.6\mu\mathrm{m}} \approx
 0.8$, $g_{1.6\mu\mathrm{m}} \approx 0.6$) as a shaded area. This
 corresponds to models that would reproduce both observed azimuthal
 brightness profiles.

 The dust grains we have used so far
 cannot account for the observed scattering properties, 
 regardless of the  maximum grain size (Fig.~\ref{fig:g_vs_g}, full line).
 Indeed, they cannot result in an asymmetry parameter
 larger than $\approx 0.65$.  Within the assumption that the
 scattering properties of the grain can be represented with the Mie
 theory, there are two ways to
 increase the scattering anisotropy: larger grain sizes and lower
 refractive indices.

Then, we first tried to see whether we could reproduce
the $g$ values at both wavelengths with a different grain size
 distribution, by varying the slope of the distribution. 
We find a solution with a slope $p = 0$ (Fig.~\ref{fig:g_vs_g}, dot-dash line),
 which is noticeably flatter than that of interstellar medium grains. This
 corresponds to a dust population almost devoid of very small grains
 ($< 0.1\,\mu$m). Models of
 dust coagulation \citep{Dullemond05,Ormel07} show that, at least
 during the first stages of the process, the slope of the grain size
 distribution does not vary much in the regime that we are interested
 in. Therefore, we consider that the solution with a flat grain size
 distribution is unlikely to occur.

The optical properties of amorphous olivine we have used correspond to compact grains. 
 An attractive solution to increase the scattering anisotropy is to
 consider porous grains, composed in a high fraction by vacuum, which
 will result in lower refractive indices. The
 porosity is defined as the fraction of the grain volume composed of vacuum:
  $ \mathcal{P} = V_\mathrm{vacuum} / V_\mathrm{grain} = 1 -  V_\mathrm{solid} / V_\mathrm{grain}$,
 where $V_\mathrm{vacuum}$, $V_\mathrm{solid}$ and $V_\mathrm{grain}$
 represent the volumes of the vacuum component, of the solid component
 (olivine silicates in this case) and the total volume of the grain,
 respectively.
We calculate the optical properties of these porous grains assuming the
 Bruggeman effective medium mixing rule \citep{Bruggeman35}.
   The dashed line in Fig.~\ref{fig:g_vs_g} represents the multi-wavelength
 scattering properties of olivine grains with a porosity  $\mathcal{P} = 0.8$.
 These grain complexes, with a maximum grain size of around $0.7\,\mu$m, mimic
 well the scattered properties inferred from {\it HST} images and in
 particular predict a more forward throwing scattering at $0.6\,\mu$m
 than at $1.6\,\mu$m (Fig.~\ref{fig:best_IMLup_scatt}, blue dotted lines). 
We consider this as an indication that we
may be observing scattered light by highly porous dust grains. 

 We also explored other dust compositions, in particular the
 porous mixture of silicates and carbon of \citet[model~A]{Mathis89}.
 This model also has a porosity $\mathcal{P} = 0.8$, and the
 scattering properties are very similar to those of our porous olivine
 dust grains. This model only predicts very faint silicate
 emission features and cannot account for the observed {\it IRS} spectrum.
 However, this reinforces our conclusion that grains with high
 porosity, independently of their exact composition, may provide a
 good explanation for the observed brightness profiles.
 Interestingly, scattering properties of porous grains have been found
 to be a good representation of those of fluffy aggregates, as long as
 the size of the inclusions is in the Rayleigh regime
 \citep{Voshchinnikov05}. If the inclusions are larger, then the
 accuracy of the effective medium theory becomes insufficient and more
 complex models, such as multi-layered spheres, are required to obtain a
 precise description of scattering properties.
 But the general trend, of an increase of
 forward throwing scattering with porosity, remains valid. 

The presence of porous aggregates suggested by the scattered
  light images is very likely to also modify the SEDs, in
  particular the profile of the silicate bands, shifting the peak 
 wavelength and affecting the shape of the bands
 \citep{Min06,Voshchinnikov08}, as well as the dust 
 emission in the millimetre regime \citep{Wright87}.
Both the opacity at long wavelength (see for
instance Fig.~4 in \citealp{Wright87}) and in the silicate bands are
however strongly dependent on the shape of the aggregates, and not only
on the degree of porosity.
\cite{Voshchinnikov08} find that small porous grains have signatures of large grains
whereas \cite{Min06} find that large aggregates composed of
small spheres have signatures of small grains. Different methods
are used to represent the aggregates in both cases (multi-layered
spheres and discrete dipole approximation, respectively) indicating that the micro-structure of the grains
is very likely to be a key element in the opacity of aggregates. 

 An alternative explanation is the presence of ice mantles
 (H$_2$O, CO, \ldots) around dust grains. Ices have refractive
 indices significantly smaller than rocks ($\approx$ 1.3 for water ice
 and $\approx$ 1.2 for CO$_2$ for instance). The dust grain
 temperature at the surface of the 
 disc, for radii larger than 100\,AU, does not exceed 70\,K, allowing
 the formation of such ice mantles. The dotted line in
 Fig.~\ref{fig:g_vs_g} presents the scattering properties of grains
 composed of water ice \citep{Irvine68}. With a maximum grain size around
 0.9\,$\mu$m,  these  grains can also reproduce the scattering
 anisotropy at both wavelengths. 

Current observations do not allow distinction between these two
solutions (porous rock grains or grains with ice mantles). As shown in
\cite{Graham07} and \cite{Fitzgerald07} in the case of the debris disc
surrounding AU~Microscopii, resolved polarimetry is a powerful tool to
solve this ambiguity. Only porous grains can produce strongly
anisotropic scattering and a high level of polarization.

%======================================================================
\section{Discussion}
\label{sec:discussion}
\subsection{A border line CTTS but with a massive disc}

IM~Lupi displays 
only a modest amount of emission-line activity, with a H$\alpha$ 
equivalent width which is known to vary from 7.5 to 21.5\,\AA\  
\citep{Batalha93}. 
Studies of H$\alpha$ emission show evidence of accretion, including variability:
\cite{Reipurth96} concluded that the H$\alpha$ feature 
shows an inverse P Cygni profile (classification IV-Rm) and
\cite{Whichmann99} observed a III-R profile.
\emph{International Ultraviolet Explorer}  low
dispersion LWP spectra show that the Mg\,II 2\,798\,\AA\ line also varies, by
a factor of about 2 in net flux \citep{Valenti03}. 
The relatively weak emission lines and lack of optical veiling 
caused \cite{Finkenzeller87} and \cite{Martin94} 
to classify this object as a weak-line T~Tauri star, although our
results show it would be 
more properly categorised as a borderline classical T Tauri star. 
IM~Lupi is a relatively weak-lined T Tauri star with 
a large and massive circumstellar disc. 
Our modelling indicates a large dust disc mass of $M_\mathrm{dust} =
10^{-3}\,M_\odot$, extending up to a radius of 400\,AU.
This large mass is puzzling given the weakness of the  H$\alpha$ line,
which suggests a low accretion rate. It is however possible that the
diagnostics of accretion have only been observed during periods of
low or moderate accretion.

\subsection{Disc structure \label{sec:disc_structure}}

\begin{figure}
  \includegraphics[width=\hsize]{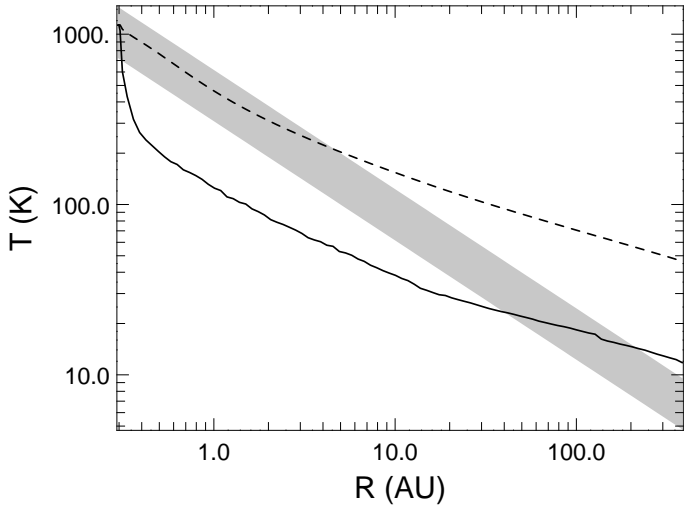}
  \caption{Temperature structure of the disc. The full and
      dashed lines represent the midplane and surface temperatures,
      respectively. The disc surface at a given radius is
      defined as the altitude above the midplane where the temperature
    is maximal. Both temperatures become equal at the inner edge of
    the disc because it is directly heated by the stellar radiation. The shaded area represents the temperature
    corresponding to the vertical gas scale height of the best model,
    assuming that the disc is vertically isothermal and that the
    central mass is between 0.5 and 1\,M$_\odot$. \label{fig:temperature}}
\end{figure}

Our multi-technique modelling allows us to quantitatively constrain
most of the geometrical parameters of the disc. A flared geometry with
a scale height of 10\,AU at a reference radius of 100\,AU is required.
The best model has a midplane temperature of 14\,K at
100\,AU. If we assume the disc is vertically isothermal (with the
temperature equal to the midplane temperature), the
hydrostatic scale height $\sqrt{k_B r^3 T(r) / G M_* \mu}$ (where
$\mu$ is the mean molecular weight) is 12 and
8.5\,AU, respectively, for a central mass object of 0.5 and
1\,M$_\odot$, respectively. This is in good agreement with the scale height
deduced from the observations (10\,AU), indicating that the outer
parts of the disc are very likely
to be in hydrostatic equilibrium. Figure~\ref{fig:temperature} presents the calculated
  midplane  temperature compared to the temperature corresponding to
  the gas
scale height of the best model. The agreement is very good at the
inner edge and in the outer parts of the disc. 
In the central parts of the disc ($<30$\,AU and excluding the very inner edge), the
midplane temperature obtained in our passive disc model is too low to
explain the scale height required by observations, indicating that
an additional heating mechanism (like viscous heating) may be at play in the
disc. 

The inner radius is constrained to be between 0.25 and 0.40\,AU, which
corresponds to a maximum dust temperature around 1\,000\,K. The
modelling of the SED alone allows values of the inner radius down to
0.15\,AU corresponding to temperatures close to the dust sublimation
temperature (1\,500\,K).  The modelling of other
observations gives additional constraints on the disc parameters and
because of the correlations between parameters, this reduces the range
of possible values for the inner radius. However, scattered images and
millimetre visibilities constrain the large scale structure of the disc. 
Because we assume that the disc can be described in terms of power-laws from
the inner edge to the outer radius, these constraints affect the
derived inner
radius. However, it is very likely that the description of the disc
using power-laws is too simplistic, and more complex geometries may
slightly shift the inner radius inwards, making it compatible with the
dust sublimation radius.

The surface density exponent is found to be close to $-1$. This value
corresponds to the median measurement for discs in Taurus-Aurigae 
obtained by \cite{Andrews07}. This also corresponds to the theoretical
value for a disc in steady-state accretion \citep{Hartmann98}.
The surface density at
5\,AU is 70\,g.cm$^{-2}$. This value is within the broad peak
around the median value of 14\,g.cm$^{-2}$ for
Taurus \citep{Andrews07}.

Considering a probable stellar mass of $\approx
1\,M_\odot$ and a gas to dust
mass ratio of 100, the disc to star mass 
ratio is $\approx$ 0.1, meaning the disc may be unstable through
gravitational collapse. The stability of the disc will depend on the
surface density.
 Our derived value of
the surface density 
exponent $\alpha$=-1, as opposed to values of 0 or -2, provides
disc stability at all radii according to the Toomre stability
criterion \citep{Toomre64}.
Collapse of gravitationally unstable discs \citep{Durisen07PPV} is one suggested mode for planet
formation. The disc of IM Lupi representing about 1/10 of the star
mass, local enhancement of density may be sufficient to start
planet formation in the disc following this process.

\cite{Hughes08} have shown that simultaneous studies of the dust
continuum and CO emissions in several well-studied discs can be reproduced with disc models
 that include a tapered exponential outer edge and not a sharp outer
 radius as we have used here.
Current observations of IM Lupi do not allow us to study in details the outer
edge of the disc but the \emph{NICMOS} image indicates that dust grains are
still present at radii larger than 400\,AU.
However, as the counter nebulae of the disc is seen in scattered light images
at 0.8 and 1.6\,$\mu$m, we can get a rough estimate of the maximum value for the
surface density in front of this second nebulae. Dust present at radii larger than 
400\,AU must be optically thin, allowing us to see the counter
nebulae through the disc and the tentative envelope.
This can be translated into a upper limit for the surface density of
the disc\footnote{The constraint is stronger at
  0.8\,$\mu\mathrm{m}$ than at 1.6\,$\mu\mathrm{m}$, due to the larger
dust opacity.}: $\Sigma (r\approx500\,AU) \times \kappa_{0.8\mu\mathrm{m}}
\lesssim 1$. If we assume that the dust composition and
grain size distribution are the same as in the rest of the disc, this
corresponds to $\Sigma (r\approx500\,AU) \lesssim
0.2\,$g.cm$^{-2}$ at 500\,AU, \emph{i.e.} about 3.5 times smaller than the
extrapolated density from our disc model, assuming it extends at radii
larger than 400\,AU. This implies that the density at radii larger
than 400\,AU must decrease significantly faster than the $1/r$
dependence we found for the disc in regions inside 400\,AU.

\subsection{Grain growth and dust settling\label{sec:dust_evolution}}

\begin{figure}
  \centering
  \includegraphics[width=0.944\hsize]{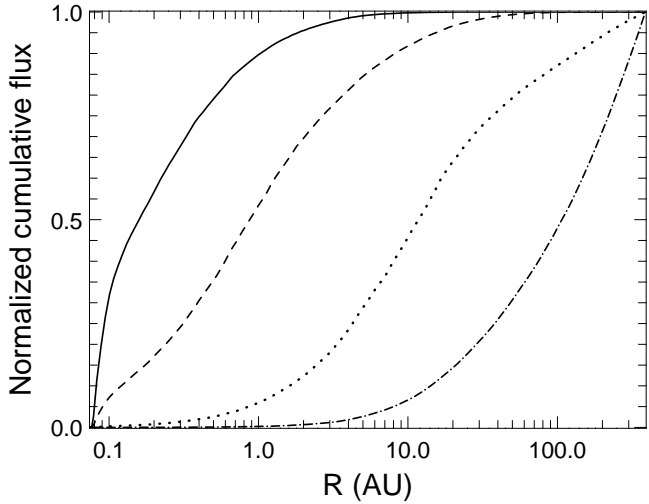}
  \caption{Cumulative received fluxes for the best model as a function
    of the distance from the star at a wavelength of 10\,$\mu$m (full
    line), 20\,$\mu$m (dashed line), 70\,$\mu$m (dotted line) and
    1.3\,mm (dot-dash line).\label{fig:flux_cumule}}
\end{figure}

As with many other classical T Tauri stars, the slope of IM Lupi's
millimetre continuum suggests a dust opacity following a law close to
$\kappa_\mathrm{abs}(\lambda) \propto \lambda^{-1}$, indicating dust
grains in the disc are larger than those in the interstellar medium.
Millimetre photometry mostly traces the thermal emission of cold dust
in the outer part ($> 15$\,AU) of the disc midplane (see
figure~\ref{fig:flux_cumule}), while the hot grains in the central
regions contribute little to this emission.

The strong silicate features in the mid-IR spectrum
indicate the presence of micron-sized grains in the disc surface
inside the first central AU. The silicate emission indeed comes from the central
parts of the disc: 
90\,\% of the emission at 10 and 20\,$\mu$m
originates from radii smaller than 1 and 10\,AU, respectively
(Figure~\ref{fig:flux_cumule}).
 Larger grains ($\gtrsim 3\,\mu$m) appear to be almost absent in these
 regions (section \ref{sec:mineralogie}). 

As already mentioned, the slope of the mm continuum and the mid-IR
silicate features indicate a
stratified structure for the disc, with large grains in the deeper
parts of the disc and small grains ($\lesssim 1\,\mu$m) in the
surface of the disc. This surface layer of small grains remains
difficult to characterise precisely. 
We have explored a structure where stratification of grain size is
vertical and caused by  settling of larger particles towards
the disc midplane. Our model with vertical settling allows us to
accurately reproduce the 
SED, scattered light images and millimetre emission maps.
However, we cannot
assess the uniqueness of the solution and other explanations may be
envisaged.
 The warm region emitting the 10 micron silicate feature 
is close to the star, while the mm/submm continuum comes from the whole
volume of the disc. 
We may be observing small particles close in and
big grains in the outer regions of the disc. However, most physical processes:
grain growth, radial drift by gas drag in the disc midplane, radiation
pressure or stellar wind, will preferentially result in larger grains
in the central regions and/or remove small grains from these regions.
Mid-infrared  interferometric observations
 (\citealp{vanBoekel04} for HAeBe stars and \citealp{Ratzka07} and \citealp{Schegerer08} for
 the T~Tauri stars TW~Hydra and RY~Tau respectively)
 have confirmed these 
trends by showing the presence of larger and more processed grains at
small spatial scales. 
Timescales for the production of large grains are significantly
shorter in the central parts of the disc and it is difficult to
imagine a physical process that will efficiently produce large grains
in the outer disc without also producing them in the inner disc. 
Even if the current observations do not allow us to firmly conclude
this point, vertical grain size stratification, 
probably resulting from a combination of settling and enhanced grain
growth close to the midplane, appears to be a
more natural explanation, and is, for now, our preferred model for the
disc of IM~Lupi.

Following the previously described argument, the presence of
millimetre grains at large radii strongly suggests that such large grains are
also present in the central AU, where the process of grain growth
should be more efficient. The 10\,$\mu$m features, dominated by the
emission from grains of size around 1.5\,$\mu$m, indicate that the
mixing in the disc is sufficient to maintain micrometric grains in the
disc surface. Moreover, the absence of 3\,$\mu$m grains in the
\emph{IRS} spectrum indicates that the decoupling between gas and dust
starts for a grain size between 1 and a few $\mu$m, \emph{i.e.} grains
larger than this threshold are settled below the
surface $\tau_{10\,\mu\mathrm{m}}$ = 1. Given the high densities in
the inner regions of the disc, an increase of 1 in optical depth is obtained
over a very small spatial scale. Even a moderate amount of dust
settling is sufficient to produce the observed effect. The very low
 values we derive for the settling index confirm that the settling remains
efficiently counterbalanced by vertical mixing, due to turbulence
for instance.

It is very likely that the process of dust settling evolves
with different timescales as a function of the distance from the star.
The silicate emission bands provide strong indications of the
efficiency of dust settling as a mean of removing grains larger than a few microns from the upper layers in the
central parts of the disc. The SED also gives some insights on the presence
 of settling in the outer parts.  
The \emph{MIPS} far-IR fluxes, which are low compared to the mid-IR
emission, indicate that the disc intercepts a relatively small fraction of the
stellar radiation at large radii ($> 10$\,AU)
compared to the fraction intercepted
at radii $< 10$\,AU probed in the mid-IR (see
Fig.~\ref{fig:flux_cumule}).
This is expected in
the case of dust settling (see for instance Fig.~7 in
\D04). The decreasing SED of IM~Lupi in the
mid-infrared is very reminiscent of the models including dust settling
of \D04. Thus, we tried to fit the SED without taking into account
the mid-infrared fluxes (between 5 and 35\,$\mu$m), and hence not
considering the silicate features. The resulting Bayesian
probabilities still exclude models without dust
settling, which do not manage to
reproduce simultaneously both the millimetre and far-infrared fluxes.
We conclude that dust settling is very likely to occur even in the
outer regions of the disc.

\D04 predicts that settled discs may become
undetectable in scattered light due to the formation of a self-shadowed
opacity structure in the outer disc. This is clearly not the case for
IM~Lupi. As noted by \D04, the self-shadowed structure
only appears for low values of the turbulence. This may indicate that
the turbulence level in IM~Lupi is large enough to prevent the disc
from self-shadowing or that we are observing the outer disc at a relatively
early stage of the dust settling process.

\subsection{Evolutionary state of IM~Lupi \& comparison with other
  classical T~Tauri stars}

Our modelling has lead to a very detailed picture of the disc
surrounding IM Lupi. We have already seen that the disc structure of
IM Lupi is similar to those of other classical T~Tauri stars.
In this section, we compare the signature of dust evolution in the
disc with the results obtained for other T~Tauri discs, in order
to determine  whether it is a singular
object or whether it can be considered as 
representative of other T~Tauri stars.

\begin{figure}
  \includegraphics[width=\hsize]{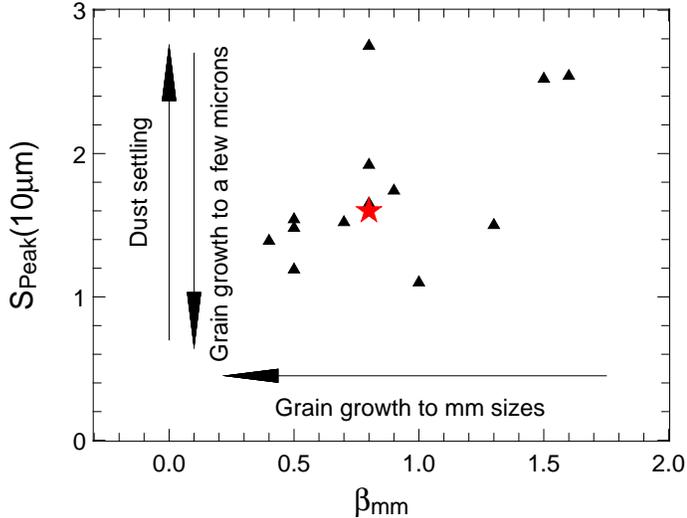}
  \caption{Strength of the 10$\,\mu$m
    silicate features as a function of the millimetre opacity slope
    for the T~Tauri stars listed in Table~\ref{tab:beta}. IM~Lupi is
    represented by the red star. The effect of grain growth and dust
    settling are schematised by arrows.\label{fig:beta}}
\end{figure}

Our conclusions on the degree of dust settling in IM~Lupi result from
the strong silicate emission feature and shallow millimetre
spectral slope. Other objects show similar characteristics. 
Thus, Fig.~\ref{fig:beta} presents the strength of the
10\,$\mu$m silicate as a function of the exponent of the opacity law
in the millimetre for a set of T~Tauri stars
(Table~\ref{tab:beta}). 
IM~Lupi is located in the bulk of T~Tauri stars and our results can
probably be extrapolated to other sources: detailed analyses of the
individual sources are required, but it is likely that at least some
of them are undergoing some dust settling. A semblable conclusion
  was reached by \cite{Furlan05,Furlan06} and \cite{D'Alessio06} based on similar
  evidence \emph{i.e.}, the presence in the SEDs of 10 and 18$\,\mu$m emission
  silicate bands and the slope and fluxes at  \mbox{(sub-)millimetre} wavelengths.

\begin{table}
  \caption{Millimetre opacity slopes and strengths of the 10$\,\mu$m
    silicate features of T~Tauri stars. References: (1) = \cite{Lommen07}, (2) =
    \cite{Kessler-Silacci06}, (3) = \cite{Przygodda03}, (4) = \cite{Rodmann06}, (5) =
    \cite{Furlan06}. The source T~Cha presented in \cite{Lommen07}
    was not selected, because of the presence of PAH emission. Both
    \cite{Rodmann06} and \cite{Lommen07} take a contribution of
    optically thick emission into account in their derivation of
    $\beta_\mathrm{mm}$.
 \label{tab:beta}}
\centering
  \begin{tabular}{lccc}
\hline
\hline
Name & $\beta_\mathrm{mm}$ & $S_\mathrm{Peak}(10\,\mu\mathrm{m})$ & Ref. \\
\hline
IM Lup   & 0.8  $\pm$    0.25   &   1.60 &   this work   \\
\hline
HT Lup   & 0.4 $\pm$     0.5	&   1.39 &   (1, 2)\\
GW Lup	 & 0.5 $\pm$	 0.5	&   1.48 &   (1, 2)\\
CR Cha   & 1.5 $\pm$	 0.6    &   2.52 &   (1, 3)\\
WW Cha   & 0.8 $\pm$	 0.8    &   1.92 &   (1, 3)\\
RU Lup   & 0.8 $\pm$	 0.5    &   1.54 &   (1, 2)\\
RY Tau   & 0.8 $\pm$	 0.1    &   2.75 &   (4, 5)\\
FT Tau   & 0.9 $\pm$	 0.3    &   1.74 &   (4, 5)\\
DG Tau   & 0.7 $\pm$	 0.1    &   1.52  &  (4, 5)\\
UZ Tau E & 0.8 $\pm$	 0.1    &   1.65 &   (4, 5)\\
DL Tau   & 1.0 $\pm$	 0.2    &   1.1 &    (4, 5)\\
CI Tau   & 1.3 $\pm$	 0.4    &   1.5 &    (4, 5)\\
DO Tau   & 0.5 $\pm$	 0.1    &   1.19 &   (4, 5)\\
GM Aur   & 1.6 $\pm$	 0.2    &   2.54 &   (4, 5)\\
\hline
  \end{tabular}
\end{table}

The tentative correlation between the strength of the silicate feature
and millimetre spectral index suggested by \cite{Lommen07} does not
appear as clear in our larger sample of T~Tauri stars. This
correlation was interpreted as an indication of fast grain growth in
both central and outer regions of the disc.
Indeed, as grains grow from
sub-micron sizes to several microns, the $10\,\mu$m feature becomes
weaker and less peaked. When they reach millimetre to centimetre
sizes, the slope of the millimetre emission becomes shallower. 
But, as we have shown, the strength of the
silicate features is strongly related to the degree of dust
settling. The stronger the settling, the smaller the apparent
(\emph{i.e.} probed in the infrared) grain size, making it difficult
to obtain a precise estimate of the actual grain sizes present in the midplane.

The detailed analysis of the silicate features indicates a small
degree of crystallisation  ($< 10\,\%$) in spite of a high mass fraction
of micrometric (hence evolved) grains. As shown by \citet[see their Fig.~9 for instance]{Schegerer06}, this is the case for several objects
and IM~Lupi is not unique on that aspect. 
\cite{Schegerer06} did not find any correlation between the
degree of crystallisation and amount of micron-sized grains.

\cite{Furlan05,Furlan06} and \cite{D'Alessio06} have estimated the degree
  of dust settling based on the colours and SEDs of a large population of
  Classical T Tauri stars in the Taurus molecular cloud. 
 They claim that to make a synthetic SED consistent with  the median SED of class II objects in Taurus, the dust to gas mass ratio in the disc
 atmosphere should be around 1\,\% (between 0.1 and 10\,\%) of the ISM ratio. In our modelling,
 the grain size distribution and the dust to gas ratio are
 continuous functions of the height above the midplane. With $\xi =
 0.05$, the dust to gas ratio starts at 1.5 times the ISM value in the disc midplane and
 reaches 10\,\%, 1\,\% and 0.1\,\% of the ISM value at
2, 3 and 6 scale heights, respectively. 
These regions roughly correspond to what is defined as the disc
surface in \cite{Furlan05,Furlan06} and \cite{D'Alessio06}.
Although not directly comparable, our estimation of
the degree of dust settling appears consistent with the calculations of
these authors.

%======================================================================
\section{Summary}
\label{sec:conclusion}

We have constructed a high quality data set of the circumstellar disc
of IM Lupi, spanning a wide range of wavelengths, from the optical to
the millimetre. All of these observations can be interpreted in the
framework of single model. A Bayesian analysis of a large grid of models
allows us to establish strong constraints on all the parameters of the
model. Although each individual observation provides valuable
information on the disc, the presented method clearly illustrates the
need for multi-wavelength and multi-technique studies in order to obtain finer
understanding of protoplanetary discs.
The conclusions of this work are:

\begin{enumerate}
\item The disc structure is very well constrained. The disc extends
  from an inner radius $< 0.4\,$AU, compatible with the dust
  sublimation radius,  up
  to 400\,AU. The scale height is 10\,AU at 100\,AU and varies with a
  flaring index of 1.15, indicating that
  the disc is in hydrostatic equilibrium in its outer parts. The slope of the surface
  density is confined to values close to $-1$, which is in agreement with the
  median value of \cite{Andrews07} and 
  predictions of steady-state accretion disc models.
\item The millimetre spectral index indicates that grains have grown
  up to a few millimetre in sizes in the disc midplane.%, at large
\item The strong  silicate emission bands, probing the surface of the
  disc in the central few AUs, also present signatures of dust
  evolution. They are dominated by grains around
  1.5\,$\mu$m but are almost devoid of grains larger than 3\,$\mu$m.
\item We conclude that a spatial stratification of the dust grains,
  depending on their size is present in the disc.
  The disc of IM Lupi has probably entered the first phases
  of planetary formation: dust grains with sizes several orders of magnitude larger
  than interstellar grains are present in the disc and dust settling is
  probably occurring, at least in the central parts but potentially also
  in the outer regions of the disc. Models without dust settling are excluded
  but the settling is constrained to a low value, suggesting that
  mixing, by turbulence for instance, remains efficient in the disc.
\item IM Lupi presents signatures of crystalline grains but with a
  low overall degree of crystallinity ($< 10\,\%$). This is in
  agreement with the results of \cite{Schegerer06} who found that
  grain growth and crystallisation occurs simultaneously in T~Tauri
  discs, although grain growth is dominant. 
\item The simultaneous analysis of the 0.6 and 1.6\,$\mu$m scattered
  light images suggests that light scattering can be more forward
  throwing at short wavelengths. Although the data are marginally
    compatible with compact silicate grains, this could indicate the presence of
  grains with low refractive indices. They may be fluffy
  aggregates (porous grains) and/or the result of
  the formation
  of ice mantles around grains. Both phenomena are
  expected to occur in discs and additional information (like
  polarisation) is required to distinguish between them.  
\end{enumerate}

The combination of our observations and modelling makes  IM~Lupi one
of the best studied 
protoplanetary disc surrounding a solar mass star.
With the exception of its low accretion signatures, all of the 
observations indicate that IM~Lupi is a typical classical T~Tauri
star. Although significant differences are expected between
individual objects, IM~Lupi can probably be considered as good prototype
of protoplanetary discs for further studies.

%======================================================================

\begin{acknowledgements}
Authors would like to thank T.~Hill for valuable comments on the manuscript.
Computations presented in this paper were performed at the Service
Commun de Calcul Intensif de l'Observatoire de Grenoble (SCCI) and on
the University of Exeter's SGI Altix ICE 8200 supercomputer. 
C. Pinte acknowledges the funding from the European Commission's Seventh Framework Program as a 
Marie Curie Intra-European Fellow (PIEF-GA-2008-220891).
The
authors thank the {\sl Programme National de Physique Stellaire}
(PNPS) and {\sl l'Action Sp\'ecifique en Simulations Num\'eriques pour 
 l'Astronomie} (ASSNA) of CNRS/INSU, France  and Agence Nationale pour la Recherche (ANR) of France under contract ANR-07-BLAN-0221, for supporting part of
this research.
This investigation was based, in part,  on observations
made with the NASA/ESA Hubble Space Telescope, obtained at the Space
Telescope Science Institute (STScI), which is operated by the
Association of Universities  for Research in Astronomy, Inc., under
NASA contract NAS 5-26555.  These observations are associated with
programs G0/7387 and GO/10177. Support for these programs was provided by NASA
through grants from STScI. This research has made use of the SIMBAD
database, operated at CDS, Strasbourg, France, and data from the Two
Micron All Sky Survey (U. Mass, IPAC/CIT) funded by NASA and
NSF. Support for this work, part of the Spitzer Postdoctoral
Fellowship Program, was provided by NASA through contracts 1224608,
1230779 and 1256316,
issued by the Jet Propulsion Laboratory, California Institute of
Technology, under NASA contract 1407.
Astrochemistry in Leiden is supported by a NWO Spinoza grant and
a NOVA grant, and by the European Research Training Network
``The Origin of Planetary Systems'' (PLANETS, contract number
HPRN-CT-2002-00308).
\end{acknowledgements}

%======================================================================
\bibliographystyle{aa}
\bibliography{biblio}

\end{document}